\documentclass[useAMS,usenatbib,fleqn]{mnras}
\usepackage{listings}
\usepackage{amsmath}
\usepackage{natbib}
\usepackage{graphicx}
\usepackage{color}
\usepackage{rotating}
\usepackage{nicefrac}
\usepackage{txfonts}
\usepackage{trivfloat}
\usepackage{subcaption}

\trivfloat{listfloat}
\usepackage{longtable}
\usepackage[percent]{overpic}
\usepackage{xspace}
\usepackage{url}
\usepackage{mathrsfs}
\usepackage{amssymb}
\usepackage{mathtools} 
\usepackage{enumitem}
\usepackage[percent]{overpic}
\usepackage{soul}
\usepackage{dsfont}

\usepackage{epsfig}
\usepackage{amsmath}
\usepackage{rotating}
\usepackage{natbib}
\usepackage{multirow}
\usepackage[noend]{algpseudocode}

\newcommand{\python}{{\sc \tt PYTHON}\xspace}
\newcommand{\clang}{{\sc \tt C}\xspace}
\newcommand{\reb}{{\sc \tt REBOUND}\xspace}
\newcommand{\rebx}{{\sc \tt REBOUNDx}\xspace}
\newcommand{\celmech}{{\sc \tt celmech}\xspace}
\newcommand{\whfast}{{\sc \tt WHFast}\xspace}
\newcommand{\ias}{{\sc \tt IAS15}\xspace}

\lstdefinestyle{customc}{
  belowcaptionskip=1\baselineskip,
  breaklines=true,
  language=C,
  showstringspaces=false,
  basicstyle=\footnotesize\ttfamily,
}

\usepackage{array}
\usepackage{appendix}
\usepackage{comment}
\usepackage{tikz}
\usetikzlibrary{shapes.geometric, arrows, fit}
\usetikzlibrary{arrows,shapes,positioning}
\usetikzlibrary{calc,decorations.markings}

\def \aap{A\&A}

\def \aj{AJ}

\def \apj{ApJ}

\def \mnras{MNRAS}

\def \pasj{Publ.~Astron.~Soc.~Japan}

\newcommand{\mer}{{\sc \tt MERCURIUS}\xspace}

\newcommand{\wh}{{\sc \tt WH}\xspace}
\newcommand{\lf}{{\sc \tt LEAPFROG}\xspace}

\newcommand{\z}{{\bf z}}
\def\myhat#1{\hat{#1}\:}
\def\mc#1{{\mathcal{#1}}}
\def\L#1{\mc{L}_{#1}\:}

\def\gsim{\;\rlap{\lower 2.5pt
 \hbox{$\sim$}}\raise 1.5pt\hbox{$>$}\;}
\def\lsim{\;\rlap{\lower 2.5pt
   \hbox{$\sim$}}\raise 1.5pt\hbox{$<$}\;}

\usepackage{varwidth}

\tikzstyle{arrow} = [draw, -latex']

\title[REBOUNDx: A Library for Additional Forces in N-body simulations]{REBOUNDx: A Library for Adding Conservative and Dissipative Forces To Otherwise Symplectic N-body Integrations}
\date{Draft: \today{}}

\author[Daniel Tamayo, Hanno Rein, Pengshuai Shi, David Hernandez]{
Daniel Tamayo$^{1}$\thanks{NHFP Sagan Fellow: dtamayo@astro.princeton.edu}, Hanno Rein$^{2,3}$, Pengshuai Shi$^{3}$, and David M. Hernandez$^{4,5,6}$ \\
$^1$ Department of Astrophysical Sciences, Princeton University, Princeton, NJ 08544\\
$^2$ Department of Physical and Environmental Sciences, University of Toronto at Scarborough, Toronto, Ontario M1C 1A4, Canada\\
$^3$ Department of Astronomy and Astrophysics, University of Toronto, Toronto, Ontario, M5S 3H4, Canada\\
$^{4}$Harvard--Smithsonian Center for Astrophysics, 60 Garden St., MS 51, Cambridge, MA 02138, USA \\
$^{5}$Physics and Kavli Institute for Astrophysics and Space Research, Massachusetts Institute of Technology, 77 Massachusetts Ave., Cambridge, MA 02139, USA\\
$^{6}$RIKEN Center for Computational Science, 7-1-26 Minatojima-minami-machi, Chuo-ku, Kobe, 650-0047 Hyogo, Japan \\
}

\vspace{0.5\baselineskip}

\begin{document}
\maketitle

\begin{abstract}
Symplectic methods, in particular the Wisdom-Holman map, have revolutionized our ability to model the long-term, conservative dynamics of planetary systems. 
However, many astrophysically important effects are dissipative.
The consequences of incorporating such forces into otherwise symplectic schemes is not always clear. 
We show that moving to a general framework of non-commutative operators (dissipative or not) clarifies many of these questions, and that several important properties of symplectic schemes carry over to the general case.
In particular, we show that explicit splitting schemes generically exploit symmetries in the applied external forces which often strongly suppress integration errors.
Furthermore, we demonstrate that so-called `symplectic correctors' (which reduce energy errors by orders of magnitude at fixed computational cost) apply equally well to weakly dissipative systems and can thus be more generally thought of as `weak splitting correctors.' 
Finally, we show that previously advocated approaches of incorporating additional forces into symplectic methods work well for dissipative forces, but give qualitatively wrong answers for conservative but velocity-dependent forces like post-Newtonian corrections.
We release \rebx, an open-source \clang library for incorporating additional effects into \reb N-body integrations, together with a convenient \python wrapper.
All effects are machine-independent and we provide a binary format that interfaces with the {\tt SimulationArchive} class in \reb to enable the sharing and reproducibility of results.
Users can add effects from a list of pre-implemented astrophysical forces, or contribute new ones.
\end{abstract}


\section{Introduction}
The long-term dynamical evolution of planetary systems remains a rich challenge for analytical investigations.
As a result, many of the advances in the last several decades have been driven by numerical studies thanks to faster computers and the development of improved algorithms for accurate and efficient numerical solutions, e.g., the chaotic evolution of Pluto \citep{Sussman88}, chaos in the inner solar system \citep{Laskar89}, and the marginal instability of Mercury over the age of the solar system \cite{Laskar09}.
In many cases, such chaos only manifests itself after millions or even billions of orbits.
This imposes strong demands on the long-term conservation properties of numerical integration methods that can be used for such studies. 

These concerns led to the development powerful symplectic integration techniques \citep[e.g.,][]{Forest83, Neri88, Forest90, Yoshida90}.
Most notably, by exploiting the near-Keplerian planetary motions, the Wisdom-Holman map \citep{Wisdom82, Kinoshita90, Wisdom91} enabled the first direct N-body integrations of our solar system over Gyr timescales.
By respecting the Hamiltonian structure of the problem, symplectic methods render the task of integration equivalent to a canonical transformation \citep[e.g.][]{Sanz92}, which enforces the conservation of several invariants of the phase space flow. 
This avoids the secular energy error growth of many of the more general methods that are not restricted to Hamiltonian systems, which can lead to unphysical collisions in long-term simulations.

Such symplectic integrators split the problem into multiple pieces that are each evolved in sequence.
Apart from the robust geometrical and numerical properties this provides, the modularity of splitting methods makes it possible to extend them for hierarchical problems \citep{Fujii07, Portegies18book} and to incorporate different physics in a single integration \citep{Pelupessy12, Portegies18}. 
This has for example been vigorously pursued in the Astrophysics Multipurpose Software Environment ({\tt AMUSE}) package \citep{Portegies18book}.

Some effects, such as bodies' higher gravitational moments, can be expressed as simple position-dependent potentials and can be trivially incorporated into symplectic schemes.
However, many crucial perturbations such as post-Newtonian corrections, tides, or radiation forces are velocity-dependent and even dissipative. 
\cite{Malhotra94} and \cite{Cordeiro96} have proposed generalizations of symplectic schemes for incorporating weak velocity-dependent forces, but their consequences in long-term integrations are not well understood. 
While some authors perform convergence tests to check the validity of their adopted stepsize using variants of the above methods \citep[e.g.][]{Zhang07}, the numerical robustness of studies where this is not done is unclear.

A generalization of symplectic integration theory would therefore be valuable to understand the source of numerical error in the above schemes, and to provide simple estimates for its magnitude to guide the selection of timestep for a given problem.

A promising direction has been offered by \cite{Galley13} and \cite{Galley14}, who have developed a generalized theory for classical mechanics that can incorporate dissipative processes in the action governing the dynamics.
One can use this formalism to construct variational integrators that generalize symplectic methods to accurately track the changing momentum and energy of the system on long timescales \citep{Tsang15}.
Potential disadvantages of this elegant framework are that the equations of motion are necessarily implicit \citep{Tsang15}, and that adding new forces becomes somewhat more complicated than with traditional methods, since one needs to derive a `nonconservative potential' using the formalism in \cite{Galley14}.

In this paper, we explore how the established theory of symplectic integration can instead be generalized in a framework of non-commutative operators, which can be applied to both conservative and dissipative systems.
Such generalized `splitting methods' have been suggested in the field of fluid mechanics as far back as \cite{Strang68}, and more recently for celestial mechanics by \cite{Mikkola98}\footnote{See also \cite{Hernandez18} for error analysis of one-step methods from the perspective of their differential equations.}.
We show that several strong results traditionally discussed in the symplectic integration literature carry over to more general and possibly dissipative splitting methods.

Readers interested in quickly using \rebx to incorporate astrophysical effects into N-body integrations could begin at Sec.\:\ref{sec:rebx} for an overview of the implementation.
For a more gradual introduction, we introduce in Sec.\:\ref{background} the notation, and review symplectic integration and the Wisdom-Holman map in an operator-centred formalism.
We then show in Sec.\:\ref{weaksplittings} how this framework can explain the energy error behaviour of such methods under weak perturbations, both conservative and dissipative.
In Sec.\:\ref{higherorder} we review how the WH map can be extended to higher order without additional force evaluations, and show that such `symplectic correctors' can also be applied to dissipative systems and are thus more general than typically considered.
In Sec.\:\ref{velforces}, we consider the general case where analytic solutions to individual steps in a split scheme are not known, and how truncation errors from numerically integrating across these individual steps interact with the overall splitting scheme errors. In particular, we show that for conservative, velocity-dependent forces like post-Newtonian corrections, the methods of \cite{Cordeiro96} and \cite{Malhotra94} give qualitatively wrong answers, and we show how to correct them.
We summarize our results and conclude in Sec.\:\ref{conclusion}.

\section{Background} \label{background}

We begin by reviewing the basic theory behind symplectic integration.
Several excellent reviews are available \citep[e.g.][]{Kinoshita90, Sanz92}, but we choose to present it in an operator-centred framework that will generalize later to dissipative systems. 
We try to keep the introduction at the level required for the discussion in the main text. 
For a more careful introduction, and to correct some sign errors in the literature, see Appendix \ref{backgroundappendix}.

\subsection{Splitting Schemes}
We consider a system of $N$ particles with positions $\bf r_i$ and velocities $\bf v_i$.
We define a differential operator $\hat{P}$, which acts on the current state of the system ${\bf z} = (\bf r, \bf v)$ to yield $\dot{\z}$, i.e., the particular set of differential equations we are trying to solve\footnote{Throughout this paper we restrict ourselves to time-independent sets of differential equations.},
\begin{align} \label{Pz}
\myhat{P}\z&: &&{\bf \dot{r}_i} = {\bf v_i} \: &&{\bf \dot{v}_i} = {\bf a_i}({\bf r}),
\end{align} 
where ${\bf a_i}$ is the $i$th particle's acceleration vector, which depends on the positions ${\bf r}$ of all the bodies.

Next, we introduce a further level of abstraction by defining a solution to Eq.\:\ref{Pz} through a corresponding operator $\mc{P}(h)$.
This ideal integrator exactly advances the state $\z$ by a timestep $h$ according to the set of differential equations $\dot{\bf z} = \hat{P} \z$.

In general, there is no closed form solution $\mathcal{P}$ (e.g., the three-body problem), but one could split the differential equations into two pieces that could each be solved trivially in isolation,
\begin{align}
\myhat{A}\z&: &&{\bf \dot{r}_i} = {\bf v_i} \: &&{\bf \dot{v}_i} = 0 \nonumber \\
\myhat{B}\z&: &&{\bf \dot{r}_i} = 0 \: &&{\bf \dot{v}_i} = \bf a_i({\bf r}),
\label{diffeq}
\end{align}
In particular, the solution $\mc{A}(h)$ keeps the velocities constant and updates the positions by $h{\bf v_i}$, while $\mc{B}(h)$ keeps the positions constant and similarly updates the velocities using constant accelerations.
This splitting works for any differential equations of the form in Eq.\:\ref{Pz}, and is thus widely applicable.
In a Hamiltonian framework, in cases where the accelerations can be derived from a position-dependent potential, it corresponds to splitting the Hamiltonian into a kinetic piece containing all the momenta, and a potential piece containing all the positions.
We therefore refer to this scheme as a kinetic-potential splitting, which was the focus of most early symplectic integrators \citep[e.g.,][]{Forest83, Neri88, Forest90, Yoshida90}.
As we will see below, better splittings for specialized problems are possible.

The idea behind splitting methods is to alternate evolution under two or more operators that can be calculated exactly and efficiently. For example, we can define a first-order splitting scheme as 
\begin{equation} 
 {\bf z}(t + h) \approx \mc{S}\mc{A}\mc{B}(h) \: {\bf z}(t) \equiv \mc{A}(h) \circ \mc{B}(h) \: {\bf z}(t), \label{1storderKT}
\end{equation}
where $\mc{B}$ advances the state $\z(t)$ by $h$ according to its half of the differential equations, and then $\mc{A}$ advances the result by $h$ with the remaining half.
Unfortunately, doing two halves of the problem in sequence is not the same as doing the full problem all at once. 
Nevertheless, if we repeatedly apply this map to get from some $\z(t_1)$ to $\z(t_2)$, at least in the limit of a vanishing timestep, the split operations become densely interleaved and this approximation will approach the real solution.


For finite timesteps, we can also see simply that how well a split scheme will correspond to the real solution should depend on the degree to which the operators $\mc{A}(h)$ and $\mc{B}(h)$ commute with one another.
We could always split each of the steps in Eq.\:\ref{1storderKT} in half and write it as $\mathcal{A}(h/2) \circ \mathcal{A}(h/2) \circ \mathcal{B}(h/2) \circ \mathcal{B}(h/2)$.
If the order in which these operators are applied didn't matter, we could switch the middle two to yield $\mathcal{A}(h/2) \circ \mathcal{B}(h/2) \circ \mathcal{A}(h/2) \circ \mathcal{B}(h/2)$.
We could then continue this split-and-switch procedure ad infinitum until the operators were again densely interleaved and we would effectively be applying both at the same time, yielding the true solution.
So we expect the errors in splitting schemes to vanish in the limits that either $h$ or the commutator of $\mc{A}$ and $\mc{B}$ go to zero.

The formula that quantifies the above two statements is the Baker-Campbell-Hausdorff (BCH) identity, which is useful in many branches of physics involving non-commuting operators like quantum mechanics. 
To state it, we have to add a final layer of abstraction and generalize beyond just how the phase space variables $\z$ change under a particular subset of the differential equations (e.g., Eq.\:\ref{diffeq}), to how an arbitrary function $g(\z)$ changes with time.
In particular, for every subset of differential equations, e.g., ${\bf \dot{z}} = \myhat{A}\z$, we can define the corresponding Lie derivative $\L{A}$ that, when acting on $g(\z)$, yields $dg/dt$ under that particular subset of differential equations.
In the simple case of $g(\z) = \z$, we get back $\L{A} \z = \dot{\z} = \myhat{A}\z$.

The BCH formula says that when we apply Eq.\:\ref{1storderKT}, we are not solving the {\it original} set of differential equations $\L{P}\z = \L{A}\z + \L{B}\z$ (Eqs.\:\ref{Pz} and \ref{diffeq}), but rather a {\it nearby} set of differential equations
\begin{equation}
\dot{\z} = \L{\mc{S}\mc{A}\mc{B}} \z = \L{P}\z + \mc{L}_{\mc{S}\mc{A}\mc{B}}^\text{err}\:\z, \label{Lerr}
\end{equation}
where the additional error term is a power series in the timestep consisting of nested commutators $[\mc{L}_A, \mc{L}_B] = \mc{L}_A\mc{L}_B - \mc{L}_A\mc{L}_B$ \citep[e.g.][]{Saha92},
\begin{equation} \label{bchflip}
\mc{L}_{\mc{S}\mc{A}\mc{B}}^\text{err} = -\frac{h}{2}[\mc{L}_A, \mc{L}_B] + \frac{h^2}{12}[\mc{L}_A - \mc{L}_B, [\mc{L}_A, \mc{L}_B]] - \mathcal{O}\left(h^3\right).
\end{equation}
In other words, if we integrate the set of differential equations in Eq.\:\ref{Lerr} numerically to high accuracy, we  match the evolution generated by the simple splitting scheme in Eq.\:\ref{1storderKT}.
We will call these differential equations that we are actually solving the `modified set of differential equations' for the splitting scheme.
Understanding the structure of this nearby problem gives insight into the numerical errors introduced by the method\footnote{\label{converge}It is important to point out that the BCH formula represents a formal, asymptotic series, which in particular is not guaranteed to converge everywhere in phase space \citep{Wisdom18}. 
In practice, as long as the adopted timestep is $\lesssim 10\%$ of the fastest timescale in the problem, this is not typically a concern.}.

As expected from the qualitative arguments above, the modified differential equations approach the true ones as $h$ approaches zero. 
We also see that if (and only if) $[\mc{L}_A, \mc{L}_B] = 0$, then there is no error and $\mc{A}(h)$ and $\mc{B}(h)$ commute, i.e., $\mc{A}(h) \circ \mc{B}(h) = \mc{B}(h) \circ \mc{A}(h)$.

By choosing appropriate steps, one can compose $\mc{A}$ and $\mc{B}$ into higher order schemes \citep[e.g.,][]{Yoshida90}.
A widely used scheme is the time-symmetric `leapfrog' method
\begin{equation} \label{P2nd}
\mc{S}\mc{A}\mc{B}\mc{A}(h) \equiv \mathcal{A}(h/2) \circ \mathcal{B}(h) \circ \mathcal{A}(h/2).
\end{equation}
Repeated application of the BCH formula shows that this particular composition cancels out error terms linear in $h$, yielding a second-order integrator with
\begin{equation} \label{P2nderr}
\mc{L}_{\mc{S}\mc{A}\mc{B}\mc{A}}^\text{err} = \frac{h^2}{24}[2 \mc{L}_B + \mc{L}_A, [\mc{L}_B, \mc{L}_A]] + \mathcal{O}\left(h^4\right).
\end{equation}
Because the scheme is time-symmetric, errors only appear at even powers of $h$.

\subsection{Symplectic Schemes} \label{secsymplectic}
The above analysis does not depend on us solving a Hamiltonian system, and is general to both conservative and dissipative systems.
However, the insight of symplectic integration is that if each of the split operators, $\mc{A}(h)$ and $\mc{B}(h)$, are exact solutions to a Hamiltonian system, then each of them conserve various Hamiltonian invariants.
This implies that {\it any} composition of them, e.g.,  $\mc{A}(h) \circ \mc{B}(h)$, must also conserve those quantities. 
In particular, such schemes will conserve the Poincar\'e invariants, the linear and angular momentum, and the infinite differentiability class order of the governing differential equations \citep{Hernandez19}, all of which are important for the accuracy of solutions \citep{Hernandez19b}. 
Having these strong conservation properties built into symplectic schemes gives them excellent long-term behaviour. 

The main subtlety, however, is that each split operator, e.g. $\mc{A}(h)$, is conserving its own Hamiltonian, e.g. $H_A$.
It is therefore not clear what Hamiltonian an overall composition scheme like the one in Eq.\:\ref{1storderKT} is conserving.
Then again, a Hamiltonian $H_A$ yields the associated set of differential equations $\dot{\z} = \L{A}\z$ through Hamilton's equations, so it should not be surprising that this answer can also be derived from the BCH formula.

One can show that one gets analogous formulae to Eqs.\:\ref{bchflip} and~\ref{P2nderr} \citep[e.g.,][]{Yoshida93},
\begin{equation} \label{H1stflip}
H_{\mc{S}\mc{A}\mc{B}}^\text{err} = \frac{h}{2}\{H_A, H_B\} + \frac{h^2}{12}\{H_A-H_B, \{H_A, H_B\}\} + \mathcal{O}(h^3),
\end{equation}
and
\begin{equation} \label{H2nd}
H_{\mc{S}\mc{A}\mc{B}\mc{A}}^\text{err} = \frac{h^2}{24}\{2H_B + H_A, \{H_B, H_A\}\} + \mathcal{O}(h^4),
\end{equation}
where curly brackets denote Poisson brackets. 
Thus, each of the error terms in the modified differential equations (e.g., Eq.\:\ref{bchflip}) can be derived from a corresponding error Hamiltonian (e.g., Eq.\:\ref{H1stflip}) \citep[e.g.,][]{Saha92}.
This is another way of seeing that the integration errors respect the symplectic geometry of the problem.

\subsection{The Wisdom-Holman Map}
While the energy errors in kinetic-potential splittings do not drift secularly with time, they remain large throughout the integration. 
Physically, this is because one is making drastic deviations from the true trajectory each timestep, i.e., force-free motion along $\mc{A}(h)$ alternating with order-unity kicks in the particle velocities under $\mc{B}(h)$.

In the case of weakly perturbed systems, like planets moving on nearly Keplerian orbits, it is much more powerful to instead split into an integrable dominant operator and the weak perturbation. 
Such a scheme, today known as the Wisdom-Holman map, proved a major development for planetary integrations, enabling the first integrations of the solar system over Gyr timescales \citep{Wisdom91}.

While the development of the Wisdom-Holman map was not driven by this operator framework (see \citealt{Wisdom18} for a historical perspective), it is instructive for our purposes to analyze it in this light.
Instead of splitting the differential equations into components of comparable magnitudes like in Eq.\:\ref{diffeq}, the Wisdom-Holman map splits them as \citep{Wisdom91, Kinoshita90},
\begin{align}
\L{K}\z&: &&{\bf \dot{r}_i} = {\bf v_i}, \: &&{\bf \dot{v}_i} = {\bf a_i^{Kep}}({\bf r}) \nonumber \\
\epsilon_p \L{I}\z&: &&{\bf \dot{r}_i} = 0, \: &&{\bf \dot{v}_i} = \epsilon_p {\bf a_i^{int}}({\bf r}) \label{diffeqWH}
\end{align}
where the ${\bf a_i^\text{Kep}}$ are the two-body Keplerian accelerations, and the ${\bf a_i^\text{int}}$ are the remaining interaction accelerations\footnote{These capture both the direct gravitational interactions between planets and their indirect effects on one another as they each pull on the central star. There are many ways to do this split between Keplerian and interaction accelerations \citep{Hernandez17,ReinTamayo2019}. The splitting above into two operators corresponds to using Jacobi coordinates \citep{Wisdom91}.}.
We have also introduced a factor of $\epsilon_p$ of order the characteristic planet-star mass ratio\footnote{This assumes that the distances from the planets to the central star are comparable to their interplanetary separations. In general, $\epsilon_p$ would capture the ratio of the perturbation accelerations to those of the dominant Keplerian motion.} to keep track of the fact that the interplanetary accelerations (and therefore the changes in the interaction steps) are much smaller than the dominant Keplerian accelerations due to the central star.

Like in the kinetic-potential split, the corresponding $\mathcal{I}(h)$ is trivial, since it keeps the positions (and thus the accelerations) constant across the timestep. The exact solution for the velocities is then just
\begin{equation} \label{vupdate}
{\bf {v}_i}(t+h) = {\bf {v}_i}(t) + \epsilon_p h {\bf a_i^{int}}.
\end{equation}
By contrast, the Kepler step $\mathcal{K}(h)$ updates both the positions and velocities of each body along their respective unperturbed two-body orbits. 
Interestingly, more accurate and efficient schemes for solving the Kepler problem have continually been found for over three hundred years \citep[e.g.,][]{Newton1687, Machin1737, Rambaut1890, Plummer1896, Brown31, Danby92, Mikkola99, WHFAST, Wisdom15}.

\subsection{Energy Error Estimates} \label{secEerr}
\begin{figure}
 \centering \resizebox{\columnwidth}{!}{\includegraphics{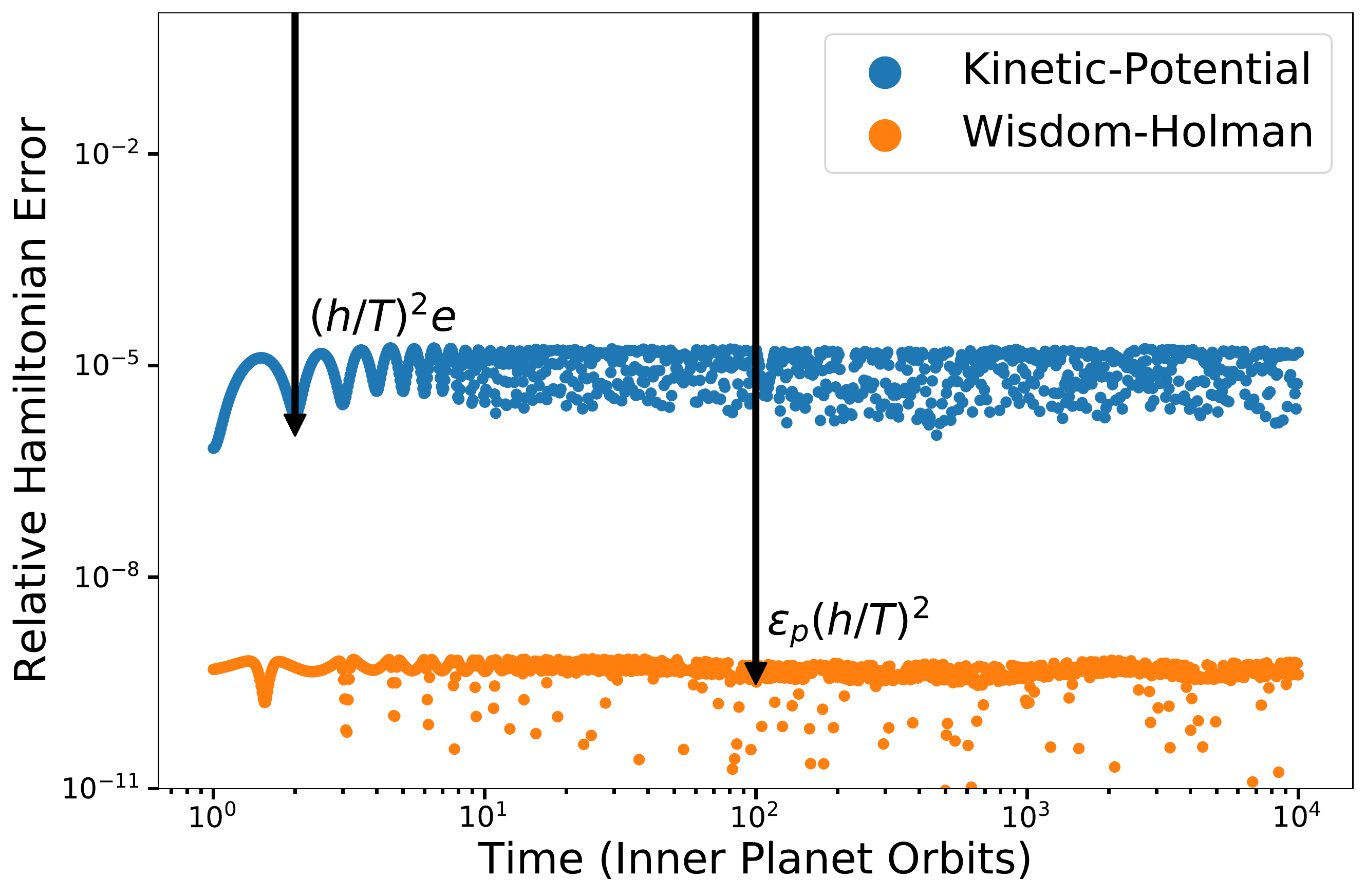}}
 \caption{Integrations of two nearly circular ($e$=0.01) Earth-mass planets around a solar-mass star with a timestep $h$ of 1\% the inner planet's orbital period $T$.
 We compare a kinetic-potential splitting to a Wisdom-Holman split. Black lines show the scheme error estimates in Sec.\:\ref{secEerr}, except the kinetic-potential splitting error has an additional factor of $e$ (Sec.\:\ref{geom}).
\label{WHvsLF}}
\end{figure}

As an example, consider a case of two Earth-mass planets with semi-major axes of 1 and 2 AU around a solar-mass star ($\epsilon_p = 3\times10^{-6}$) with orbital eccentricities of 0.01.
We specialize to second order schemes (Eq.\:\ref{P2nd}), and compare the kinetic-potential splitting (Eq.\:\ref{diffeq}) to the Wisdom-Holman splitting (Eq.\:\ref{diffeqWH}).
Figure \ref{WHvsLF} plots both methods' fractional energy errors over ten thousand inner-planet orbits, using a timestep of 1\% the innermost planet's orbital period.

The Wisdom-Holman map conserves the energy many orders of magnitude better than the kinetic-potential splitting, and we can make simple order of magnitude estimates to understand why.
This is clearest in a Hamiltonian framework for conservative systems \citep[e.g.][]{Wisdom91, Saha92}, but we instead show how it arises in the above operator-centred picture that will later generalize to dissipative systems.
While this is not a rigorous derivation, it is useful to be able to quickly estimate expected errors and their associated scalings.
While we do not plot all possible dependencies, we have verified that the various scalings predicted below are correct.

A useful feature of the abstract Lie derivative formalism of Sec.\:\ref{background} is that they describe not only how the phase space variables $\z$ evolve, but also how functions like the energy $E(\z)$, change with time.
In particular, just like $\dot{\z}^{\rm err} = \mc{L}_S^{\rm err}\z$ yields the additional terms in the modified differential equations for the phase space variables using a particular split scheme $\mc{S}$ (Sec.\:\ref{background}), $\dot{E}^{\rm err} = \mc{L}_S^{\rm err}E(\z)$ yields the additional terms in the modified differential equations for the energy evolution.

To approximately estimate the error $\Delta E^{\rm err}$ they cause, we can multiply the additional terms $\dot{E}^{\rm err}$ by the timescale over which these errors coherently add up.
Toward this end, on the left of Fig.\:\ref{WHvsLF}, the logarithmic time axis makes it possible to see that the errors accumulate coherently over several timesteps ($h=0.01$), and that on longer timescales, they oscillate.
We can understand this from Sec.\:\ref{background}.
Because the BCH formula says that all the error terms in the modified differential equations are composed of the split operators that propagate the real dynamics (Eq.\:\ref{P2nderr}), the errors will inherit the timescales in the original problem.
In particular, the timescale over which errors will accumulate coherently will be the shortest timescale in the problem.
In this case of near-circular, non-resonant orbits, $T$ would be the synodic period on which planets kick each others' orbits every conjunction.
In our typical example where the orbits are not too close, this is approximately the orbital period of the innermost planet; for a very eccentric orbit, $T$ would be the much shorter timescale for pericentre passage \citep[e.g.,][]{Rauch99, ReinTamayo2015}, etc.
We thus have
\begin{equation} \label{Eerr}
\Bigg|\frac{\Delta E^{\rm err}}{E}\Bigg| \sim \Bigg|\frac{T \dot{E}^{\rm err}}{E}\Bigg| = \Bigg|\frac{T\mc{L}_S^{\rm err}E}{E}\Bigg|,
\end{equation}
where, for the second-order schemes used in Fig.\:\ref{WHvsLF}, $\mc{L}_S^{\rm err}$ is given by Eq.\:\ref{P2nderr}.

Since $\L{A}$ and $\L{B}$ each yield time derivatives according to their own subset of the differential equations, we can make a rough estimate in Eq.\:\ref{Eerr} by replacing each instance of $\L{A}$ and $\L{B}$ in $\mc{L}_S^{\rm err}$ by their respective inverse characteristic timescale. 
In this case we are referring specifically to the timescale on which each operator {\it in isolation} would change the particle states by order unity, i.e., $\text{z}/\dot{\text{z}}$.

For Kepler splittings, the Kepler Lie derivative $\L{K}$ induces order-unity changes on the orbital timescale\footnote{More precisely $2\pi/T$. Considering the simple 1-D case of a circular orbit with semimajor axis $a$ and orbital frequency $\omega$, $\dot{\text{v}} = \omega^2 a$, so $\L{K} \sim \text{v}/\dot{\text{v}} = 1/\omega$. We are ignoring these factors of $2\pi$, which largely cancel with the coefficients in the BCH formula, but they can be important for higher-order schemes that accumulate many such factors.} 
 $T$. Perturbation operators like the interaction step in the WH map induce order-unity changes on timescales that are longer by roughly a factor of the perturbation strength $\epsilon$ (e.g., for the interaction operator, $\text{v}/\dot{\text{v}}_I = (\text{v}/\dot{\text{v}}_{K}) (\dot{\text{v}}_K/\dot{\text{v}}_I) = \epsilon_p T$, see Eq.\:\ref{diffeqWH}).
 
If we further assume that the two operators $\L{A}$ and $\L{B}$ do not commute, i.e., $[\L{A},\L{B}] \sim \L{A}\L{B}$, then plugging Eq.\:\ref{P2nderr} into Eq.\:\ref{Eerr} yields a fractional energy error estimate of $\sim \epsilon_p (h/T)^2$.
This gives the right scaling for a second-order method, as expected.
Indeed, the same argument applied to any Kepler splitting says that if the perturbation accelerations are a factor of $\epsilon$ smaller than the Keplerian ones, then the second-order splitting scheme given by Eq.\:\ref{P2nd} will yield energy errors $\sim \epsilon (h/T)^2$.

By contrast, in the kinetic-potential splitting (Eq.\:\ref{diffeq}), both $\L{A}$ and $\L{B}$ induce order-unity changes on the orbital timescale $T$, so $\L{A} \sim \L{B} \sim 1/T$. This yields a correspondingly larger error $(h/T)^2$.

More physically, the WH method is benefiting from the fact that the dominant motion is known and solved exactly. 
In the limit of $\epsilon_p \rightarrow 0$, the Wisdom-Holman map would yield an exact trajectory, while the kinetic-potential splitting would still be splitting the Kepler problem and the errors would remain large and unchanged.

These simple estimates (black arrow in Fig.\:\ref{WHvsLF}) match well to integrations with the Wisdom-Holman map, while our estimate for the kinetic-potential splitting is depressed by approximately a factor of $e$ from our simple estimate.
This has to do with our simplification of the commutation relations, and has interesting broader implications that we discuss in Sec.\:\ref{geom}.

In summary, while the Kepler step is more expensive to compute than the steps in a kinetic-potential splitting, the factor of $\epsilon_p$ gain makes reaching a given level of accuracy significantly faster with the Wisdom-Holman map.
Perhaps more importantly, in cases where inter-planetary forces are very weak, the Wisdom-Holman method keeps the additional terms in the modified differential equations smaller than the interplanetary terms.
By contrast, in the kinetic-potential splitting, these additional error terms can easily be larger than the inter-planetary forces, leading to spurious results.

These scalings suggest the cases where splitting methods will be most useful, i.e. when perturbations are weak $\epsilon \ll 1$ and when the shortest characteristic timescale $T$ is not too short.
For example, for a planet experiencing dynamical tides on an extremely eccentric orbit, the strength of the perturbation changes by orders of magnitude over an orbital period. 
In this case, the shortest timescale would be the characteristic timescale of pericentre passage, which can be much shorter than an orbital period.
Because simple symplectic schemes require a fixed timestep, the small $h$ required where the forces are strongest acts as a bottleneck for the remaining orbit that could otherwise be easily integrated.
In such cases it is typically more efficient to use a more general scheme with adaptive timesteps like \ias, or in some restricted cases it is possible to symplectically adapt timesteps in inverse proportion to the strength of an external potential \citep{Preto99, Mikkola99reg, Petit19}.

\section{Additional Forces} \label{weaksplittings}
We now move beyond point-particle gravity to consider adding additional effects in N-body integrations.
We will see that the above framework, by not specializing to Hamiltonian systems, can be used to straightforwardly understand the numerical behavior under both conservative and dissipative forces.

\subsection{Conservative Forces} \label{posforces}
We begin by considering conservative forces that can be derived from position-dependent potentials.
Hamiltonian perturbations that cannot be written strictly in terms of the particle positions are also important (e.g., post-Newtonian corrections), but we defer their discussion to Sec.\:\ref{velforces}.

We note that while any position-dependent potential can obviously be trivially incorporated into a kinetic-potential split, it can also be directly inserted into the interaction step in the Wisdom-Holman map as an additional position-dependent acceleration in Eq.\:\ref{diffeqWH}.
Then Eq.\:\ref{vupdate} remains the exact solution with the acceleration given by the sum of all the position-dependent accelerations in the problem.

As an example, we consider the same two-planet case as above, but now with additional perturbations from an oblate primary, with the planets orbiting in the primary's equatorial plane.
The first corrections to point source gravity in a multipole expansion of the primary's potential is the quadrupole term, with a potential in the equatorial plane of
\begin{equation}
V_{J2}(r) = -\frac{1}{2}J_2\Bigg(\frac{R}{r}\Bigg)^2 \frac{GM}{r} \equiv \epsilon_{J2} \frac{GM}{a_0}\Bigg(\frac{a_0}{r}\Bigg)^3. \label{HJ2}
\end{equation}
The innermost body will be most affected, so we take $a_0$ to be its original, reference semimajor axis, $G$ is the gravitational constant, $M$ and $R$ are the primary's mass and radius, and $J_2$ is the standard dimensionless coefficient for the quadrupole field.
We have also introduced a parameter $\epsilon_{J2} = \frac{1}{2}J_2\left(\frac{R}{a_0}\right)^2$ that captures the smallness of the effect relative to the dominant Keplerian potential $\approx GM/a_0$, since $a_0/r$ is approximately unity for nearly circular orbits.
We take $\epsilon_{J2} = 10^{-3}$.

\begin{figure}
 \centering \resizebox{\columnwidth}{!}{\includegraphics{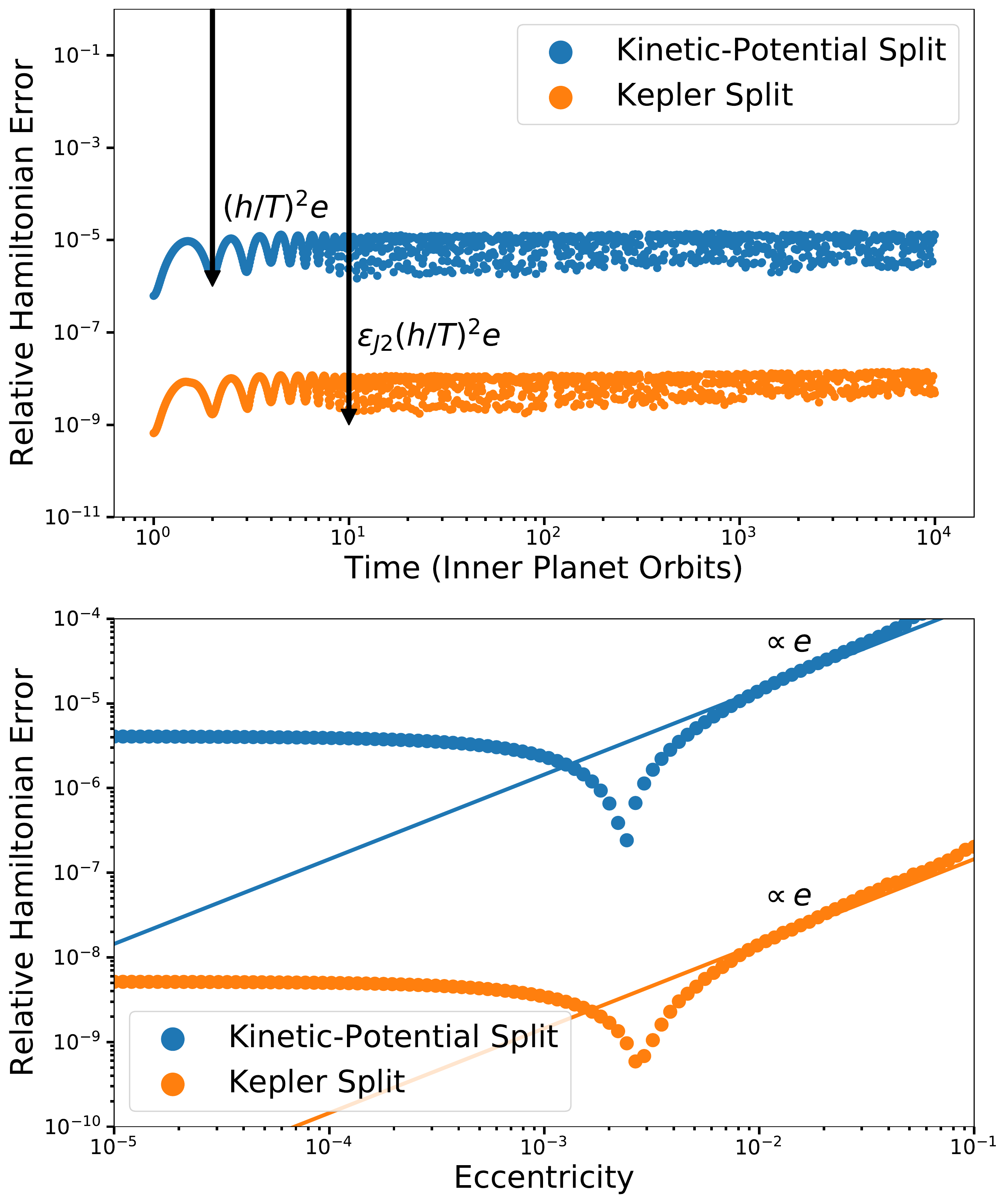}}
 \caption{Top panel is the same as Fig.\:\ref{WHvsLF}, except we now additionally include a stellar oblateness perturbation of $\epsilon_{J2} = 10^{-3}$. Estimated scheme errors in the main text are labeled and denoted by black arrows. Bottom panel shows how the fractional energy error depends on the orbital eccentricities of the planets. It is linear in the eccentricity (solid lines) down to an eccentricity $e \sim \epsilon_{J2}$, as explained in the main text.
\label{figJ2}}
\end{figure}

We again compare second-order schemes (Eq.\:\ref{P2nd}) using kinetic-potential and Wisdom-Holman Kepler splittings in Fig.\:\ref{figJ2}.
Comparing with Fig.\:\ref{WHvsLF}, we see that the kinetic-potential errors have not changed.
This is because the planetary accelerations in $\L{B}$ for the kinetic-potential splitting (Eq.\:\ref{diffeq}) are dominated by those due to the star.
Adding a small acceleration due to the oblateness perturbation therefore does not noticeably change the dominant error terms in Eq.\:\ref{P2nderr}.
By contrast, for the Kepler splitting, our chosen oblateness perturbations are much larger than the interplanetary ones ($\epsilon_{J2} \gg \epsilon_p$), so they change the error behavior.
The same argument following Eq.\:\ref{Eerr} suggests an energy error $\sim \epsilon_{J2}(h/T)^2$.

However, both our error estimates for the kinetic-potential and Kepler splitting fall short by a factor of the eccentricity (both black arrows in Fig.\:\ref{figJ2} include this factor).
This can be a significant suppression in the errors and is due to the geometrical properties of splitting methods, as we explore in the next section.

\subsection{Splitting Methods Exploit Symmetries} \label{geom}
Because for the remainder of the paper we focus on Kepler splittings, we consider in detail the eccentricity factor in the Wisdom-Holman errors seen in Fig.\:\ref{figJ2}.

The Kepler problem is special in its degeneracies; it conserves all orbital elements, except for the mean longitude $\lambda$, which advances linearly.
This implies that when alternating Kepler steps with perturbation steps, only perturbation components that change the particles' semi-major axes (which determine the mean motions $d\lambda/dt$) do not commute. 

For example, for a perturbation that only changed the eccentricity vector's magnitude and direction, it would not matter whether the Kepler or perturbation step acted first, since the Kepler step leaves those quantities unchanged.
The ordering only matters for perturbations that change the semimajor axis; if the perturbation acts first, the mean motion will change, and when the Kepler step operates, the particle will end up at a different mean longitude than if the Kepler step had acted first with the original value of the mean motion.

We can then see that for any purely radial force like the oblateness perturbation, the change in orbital energy (and thus semi-major axis)  $-{\bf F} \cdot {\bf v}$ vanishes for circular motion. 
If we slowly increase the initial eccentricity, the perturbation will add and extract orbital energy as each body moves toward and away from pericentre, to leading order in proportion to $e$. 
So our mistake in estimating the energy error in Sec.\:\ref{secEerr} was in assuming that the commutators introduced factors of order unity, i.e., that $[\L{A},\L{B}] \sim \L{A}\L{B}$.
Our physical argument says that for radially perturbed, nearly circular orbits, the operators in fact approximately commute.
The commutator $[\L{A},\L{B}]$ oscillates with an amplitude $\sim e\L{A}\L{B}$ and vanishes at pericentre and apocentre where ${\bf F} \cdot {\bf v} = 0$ and the semimajor axis is unchanged.\footnote{Actually it would not completely vanish in this problem due to the much weaker perturbations from the additional planet that would at these points become the dominant effect.}
This is physically what is causing the oscillations visible on the left of the top panel in Fig.\:\ref{figJ2}, where the logarithmic time axis renders the orbital timescale visible. 
The planets were both started at pericentre, and one can see that the energy errors remain constant at pericentre (every time unit) and apocentre (every half-time unit).
We note that the flat errors for $e < 10^{-3}$ in the bottom plot of Fig.\:\ref{figJ2} are an artefact of our numerical setup\footnote{The physical argument is correct for all {\it geometric} eccentricities (i.e., the value measured by looking at the shape of the physical orbit in space, which includes the effects of the perturbation) but the typical subtlety arises when dealing with {\it osculating} eccentricities (i.e. the transformation from positions and velocities to unperturbed 2-body Kepler orbits, which ignores the effect of the perturbation) smaller than the size of the perturbation $\epsilon$.
Because for convenience we initialize our orbits using osculating elements, decreasing the initial osculating eccentricities below $\epsilon_{J2}$ does not make them any more circular in geometric space, because the oblateness perturbations kick them away from Keplerian motion at order $\epsilon_{J2}$.
So in our integrations, the geometric eccentricity that goes into the argument above reaches a floor at $\sim \epsilon_{J2}$, with a corresponding error $\epsilon_{J2}(h/T)^2e = \epsilon_{J2}^2(h/T)^2$, which matches the flat regime on the left of the bottom panel of Fig.\:\ref{figJ2}.}.

The above example highlights the power of splitting methods' geometrical properties. 
Because all the error terms in the modified differential equations (e.g., Eq.\:\ref{P2nderr}) consist of nested commutators of $\L{A}$ and $\L{B}$, any symmetries (or near-symmetries) are inherited in the integrator's error properties to {\it all} orders. 
This means any higher-order splitting scheme that tries to correct for higher and higher order terms in the BCH expansion \citep[e.g.,][]{Wisdom96, Laskar01} will always have such a geometrical suppression in its leading error term.

This is not typically emphasized in the literature on symplectic N-body integration, presumably because the Kepler and interaction steps do not have any such symmetries (there are always conjunctions between planets that change the semi-major axes at leading order).
However, many astrophysically important effects are highly symmetric potentials, e.g. multipole gravitational potentials, radiation pressure, simple general relativistic corrections, etc.
This fact, combined with the the highly degenerate Kepler problem, often lead to strongly suppressed errors for the nearly circular or nearly coplanar orbits we often want to model.

In summary, splitting methods are often powerful not only for their long-term conservation of important quantities, but also because they can yield orders of magnitude higher accuracy for a fixed timestep.
In the above case, a specialized second-order scheme split into separate Kepler and perturbation steps reduced the errors by a factor of $\epsilon_{J2}e = 10^{-5}$ over short timescales compared to what one would obtain with a generic second-order Runge-Kutta scheme, even more over long timescales.

\subsection{Dissipative Forces} \label{secdissipation}
While dissipative systems are qualitatively different from the conservative cases above, we now show that we can similarly understand their error behaviour using the above framework.
As should be clear from the operator-centred development throughout the paper, we could take any differential equations, split them in any way we please, and if we could find time-evolution operators for those subproblems, then we could compose them and find their error behavior through the BCH formula.
In particular, the error estimates we developed in Sec.\:\ref{secEerr} and applied in Sec.\:\ref{posforces} come from the modified differential equations, without reference to conservative or dissipative forces.

On the other hand, one important distinction for symplectic systems is that the fact that the composition of two Hamiltonian operators must be Hamiltonian implies that all the BCH error terms are also Hamiltonian (Sec.\:\ref{secsymplectic}).
By contrast, when the perturbation is dissipative, the presence of $\L{B}$ operators in the error terms of the modified differential equations (e.g., Eq.\:\ref{P2nderr}) shows that there is additional damping (or injection of energy) introduced by the splitting scheme.
This generically leads to secular drifts, as we show in the following example.

Consider a single planet in orbit around its primary with an initial orbital eccentricity of $0.1$ and an orbital period of 1 year, subject to a simple damping force directed opposite to the planet's velocity vector ${\bf v}$,
\begin{equation}
{\bf F} = -m \frac{{\bf v}}{2\tau_a}, \label{Fdrag}
\end{equation}
where $m$ is the planetary mass. This parametrized prescription orbit-averages to yield inward migration with the semi-major axis decaying exponentially on an e-folding timescale $\tau_a$ \citep{Papa00}. 
We set $\tau_a = 1000$ years, and integrate for three damping timescales $\tau_a$.
In this time, the planet's semi-major axis moves inward by a factor of $\text{exp}(3) \approx 20$, and the orbital period decreases by a factor of $\text{exp}(9/2) \approx 90$.

We integrate the system using a second-order Kepler splitting 
\begin{equation}
\mathcal{K}(h/2) \circ \mathcal{D}(h) \circ \mathcal{K}(h/2),
\end{equation}
where $\mathcal{K}$ evolves the planet on a Kepler orbit, and $\mathcal{D}$ damps the motion according to Eq.\:\ref{Fdrag}. 
We adopt a timestep of $10^{-3}$ times the planet's initial orbital period, and choose to approximate $\mathcal{D}(h)$ by integrating across the timestep using a fourth-order Runge-Kutta scheme (see Sec.\:\ref{velforces}). 

\begin{figure}
 \centering \resizebox{\columnwidth}{!}{\includegraphics{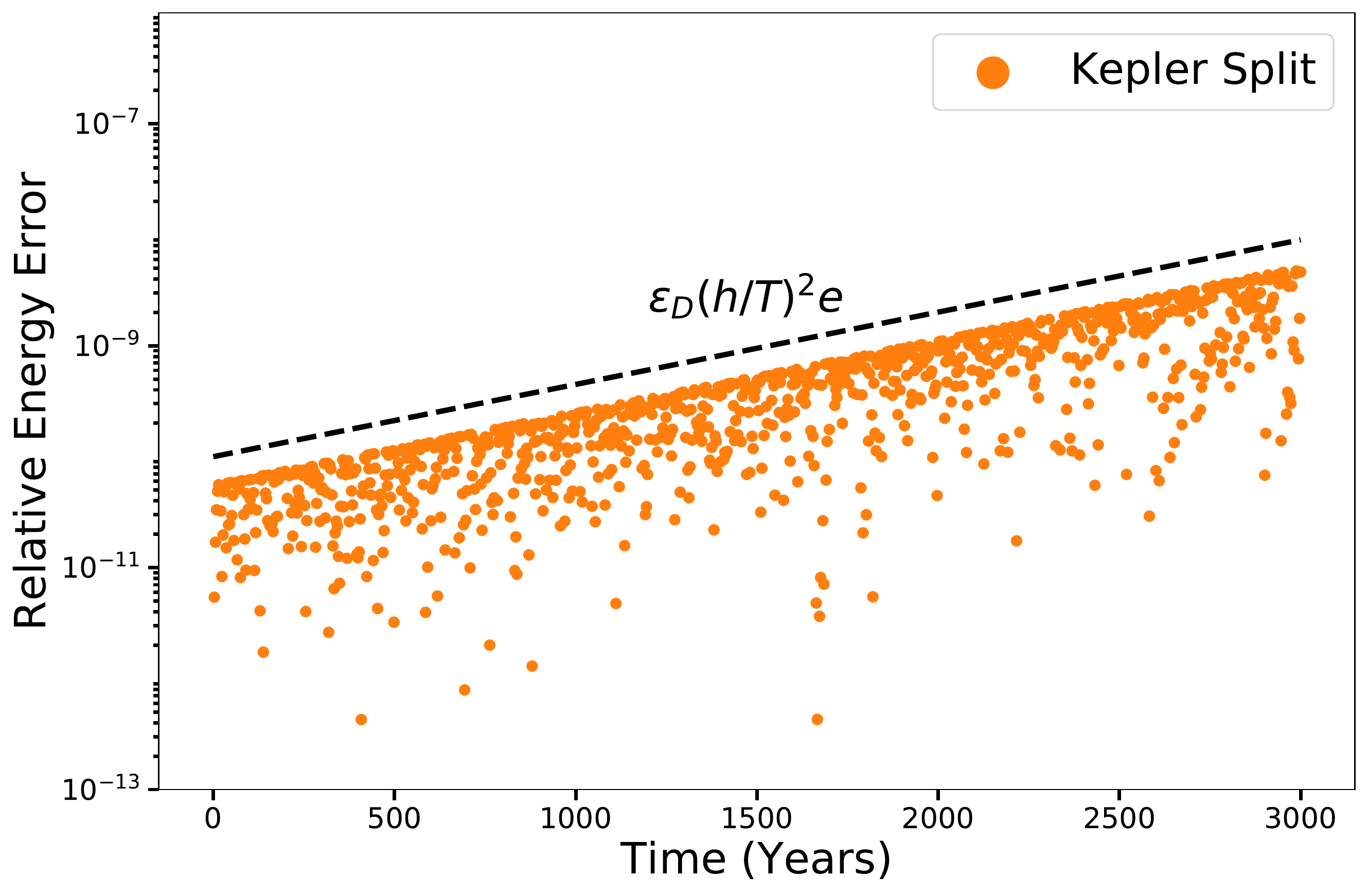}}
 \caption{Second-order, Kepler-splitting integration of a single planet initially on a 1-year, 0.1 eccentricity orbit, subject to a damping force that causes the orbital period to decay exponentially. The relative energy error follows the same estimate (dashed black line) as the conservative case in Fig.\:\ref{figJ2}, accounting for the fact that the orbital period $T$ is changing. 
\label{damping}}
\end{figure}

With dissipation, we no longer have a conserved quantity to track. 
Instead, as a proxy for the exact propagator $\mathcal{P}$, we also integrate the system with \ias \citep{IAS15}, a high-order adaptive-timestep method whose accuracy reaches machine precision.
We plot in orange in Fig.\:\ref{damping} the relative error between the energy calculated in our Kepler splitting integration and the one with \ias. 

One can see a clear secular drift. 
While the errors in the symplectic integrations in Fig.\:\ref{figJ2} oscillate and average out to yield flat time evolution, dissipative errors systematically overdamp or underdamp.

This scaling can be understood in the same way we analyzed the conservative case.
The second-order splitting scheme error estimate is (Sec.\:\ref{secEerr}) $\epsilon_D (h/T)^2$, with $T$ the planet's orbital period, and $\epsilon_D = T/\tau_a$.
However, like in the case of $J_2$ perturbations discussed above, there is additionally a geometric suppression of the error by a factor of the eccentricity\footnote{To see this we have to go one step further than the arguments in Sec.\:\ref{geom}.
There we argued that only the components of the perturbation that change the particles' semi-major axes don't commute with the Kepler step. 
Here the force is always pointed exactly opposite the particle's velocity, so the two steps definitely do not commute.  
But now we additionally have to ask whether the order of operations matters for the quantity whose error we are measuring, i.e., the energy.
For a circular orbit, it would not.
Whether or not the Kepler step moves the planet along the circle before or after the step does not affect the energy loss, since  the problem is azimuthally symmetric.
So again the energy errors are suppressed by a factor of $e$, though in this case there would be no suppression of the phase errors. 
We consider phase errors more carefully in \cite{ReinBrownTamayo19}.}.
We show this estimate in Fig.\:\ref{damping} with a black dashed line.

The secular rise then simply reflects the fact that the orbital period is changing exponentially as the planet migrates inward, which yields a straight line on the linear-log scale.

Remarkably, this suggests that exponential outward migration would yield errors that exponentially {\it decrease} with time. 
We find that this is indeed the case, though this case is more subtle.
In the case plotted above where the errors committed each timestep increase with time, the most recent errors always dominate the error budget.
In the case where the instantaneous errors are getting smaller, one might expect the total error to remain at the level incurred at the beginning of the integration, where errors are largest.
But these errors are oscillatory, and as long as the errors are changing adiabatically (i.e., as long as there are many orbits per migration timescale), the oscillations can march toward smaller amplitude.
Of course for exponential outward migration, it does not take many migration timescales until the orbital period becomes comparable to the migration timescale, at which point the applied force in Eq.\:\ref{Fdrag} (unphysically) becomes comparable to the central gravitational force, and we see numerical errors rise again.

In summary, the error behaviour in weakly dissipative splitting schemes can be understood in the same way as conservative cases, starting from the modified differential equations yielded by the BCH formula.
Dissipative splitting  schemes retain strong geometric properties (suppressing energy errors by a factor of $e$ above), and conservation properties.
For example, a dissipative perturbation that damps the radial component of particle velocities (and thus acts radially) would conserve angular momentum.
Composing such a step with conservative Kepler and interaction steps would still conserve the total angular momentum to machine precision (since each step does individually).

While we have seen that dissipative splitting schemes can systematically over(under)damp, we show in the following section that the same techniques from symplectic integration allow us to correct these errors at no additional computational cost.

\section{Higher Accuracy Splitting Methods At Fixed Computational Cost}
\label{higherorder}
\subsection{Symplectic Correctors} \label{sympcorr}

Ignoring any improvements resulting from any symmetries in the problem (Sec.\:\ref{geom}), the splitting errors for a second-order Wisdom-Holman scheme should oscillate on orbital timescales with an amplitude of $\sim \epsilon (h/T)^2$. 
However, these high frequency error oscillations should be unimportant to the long-term evolution through the averaging principle \citep{Wisdom91}.

\cite{Wisdom96} go further and show that these high-frequency oscillations of $\mathcal{O}(\epsilon)$ can be efficiently removed through a near-identity canonical transformation, which they term symplectic correctors, from the real action-angle variables to mapping variables.
In this picture, the short term oscillations intuitively arise from a mismatch in initial conditions.
When going from the real system to the modified mapping, one has to correct the initial conditions to modified `mapping coordinates'.
An integration is then of the form
\begin{equation}
\mathcal{C}^{-1}(h) \circ \mc{S}\mc{A}\mc{B}\mc{A}(h) \circ ... \circ \mc{S}\mc{A}\mc{B}\mc{A}(h) \circ \mathcal{C}(h), \label{itercorr}
\end{equation}
where each $\mc{S}\mc{A}\mc{B}\mc{A}(h)$ represents a timestep of the second-order Wisdom-Holman map (Eq.\:\ref{P2nd}), and $\mathcal{C}(h)$ and $\mathcal{C}^{-1}(h)$ are the symplectic corrector transformations to and from mapping variables at the beginning and end of the integration. \cite{Wisdom96} furthermore prove that this canonical transformation removes all the error terms in the modified differential equations\footnote{We note that ordering matters once we specialize to systems where one operator is dominant. \cite{Wisdom96} (with corrections in \citealt{Wisdom06}) give correctors for the $\mathcal{A}(h/2) \circ \epsilon \mathcal{B}(h) \circ \mathcal{A}(h/2)$ second-order scheme. The correctors for the alternative second-order scheme $\epsilon \mathcal{B}(h/2) \circ \mathcal{A}(h) \circ \epsilon \mathcal{B}(h/2)$ would have different coefficients.} for the Wisdom-Holman map (Eq.\:\ref{P2nderr}) of $\mathcal{O}(\epsilon h^n)$ for all $n$. 
The leading error using symplectic correctors is then $\mathcal{O}(\epsilon^2 h^2)$.

For generality and convenience of application, \cite{Wisdom96} go on to show how the correctors $\mc{C}(h)$ for an arbitrary pair of integration steps $\mathcal{A}$ and $\epsilon\mathcal{B}$ (hereafter we assume that $\mathcal{A}$ is the dominant operator and $\mathcal{B}$ is a perturbation) can be approximated to progressively higher order through compositions of $\mathcal{A}$ and $\epsilon\mathcal{B}$ with carefully chosen timesteps forward and backward in time. 
\cite{Wisdom06} gives explicit coefficients for such $n$th order approximations $\mathcal{C}_n$, which 
remove error terms $\mathcal{O}(\epsilon h^k)$ up to $k=n$ in Eq.\:\ref{P2nderr}.
Depending on the problem, this leaves as the leading error the larger of the first uncorrected term $\mathcal{O}(\epsilon h^{n+2})$ and the first term at higher order in the perturbation parameter $\mathcal{O}(\epsilon^2 h^2)$. 

We also briefly note that these correctors for the second-order Wisdom-Holman map $\mc{S}\mc{A}\mc{B}\mc{A}$ (Eq.\:\ref{P2nd}) can also be used for the first-order scheme $\mc{S}\mc{A}\mc{B}$ (Eq.\:\ref{1storderKT}) with a simple modification.
As far as the first-order scheme is concerned, the second-order scheme is a reasonable approximation to the exact propagator $\mathcal{P}$, so
\begin{eqnarray} \label{expandP2nd}
\mathcal{P}(h) &\approx& \mathcal{A}(h/2) \circ \mathcal{B}(h) \circ \mathcal{A}(h/2) \nonumber \\
&=& \mathcal{A}(-h/2) \circ \mathcal{A}(h) \circ \mathcal{B}(h) \circ \mathcal{A}(h/2). \nonumber \\
&\equiv& \mathcal{C}_1^{-1}(h) \circ \mc{S}\mc{A}\mc{B}(h) \circ \mathcal{C}_1(h), \label{1stordercorr}
\end{eqnarray}
and thus $\mc{C}_1(h) = \mathcal{A}(h/2)$ acts as a first-order corrector for the first-order splitting scheme to yield $\mc{S}\mc{A}\mc{B}\mc{A}$.
Then, if desired, one could use a higher order correctors $\mc{C}_n$ in Eq.\:\ref{itercorr}.
This trick of combining half steps of the second-order scheme in the middle of the integration is used widely (in \reb this is referred to as synchronization), but rarely described as a corrector that can be extended to higher order\footnote{{\bf Jack Wisdom realized this long before we did (personal communication).}}.

Finally, the fact that correctors compensate for the leading errors provides a straightforward way to interpret what the scheme is getting wrong. 
For example, inverting Eq.\:\ref{1stordercorr} yields
\begin{equation} \label{1stMIS}
\mc{S}\mc{A}\mc{B}(h) \approx \mathcal{A}(h/2) \circ \mathcal{P}(h) \circ \mathcal{A}(-h/2),
\end{equation}
i.e., to leading order, the first-order scheme $\mc{A}(h) \circ \mc{B}(h)$ is equivalent to performing an exact step  along the true solution $\mathcal{P}(h)$, except with the mistake of taking additional forward and backward Kepler half-steps before and after the fact.

\subsection{Symplectic Correctors Are Splitting Correctors} \label{seccorrectors}

While \cite{Wisdom96} derived symplectic correctors through physically motivated canonical perturbation theory specific to Hamiltonian systems, they point out that the formulas could also have been derived solely from the Lie algebra (commutators) of the two operators using Lie series and the BCH formula.
But as mentioned above, the BCH formula is a statement about the composition of non-commutative operators and has nothing to do with symplecticity. 

`Symplectic' correctors therefore should correct {\it any} second-order splitting scheme involving a dominant operator and a perturbation. 
We now demonstrate explicitly that this is true even for dissipative perturbations.
`Symplectic' correctors are therefore more widely applicable than their name implies.
Specifically, they represent `weak splitting correctors.'
In fact, like \cite{Wisdom96}, we rediscovered that similar ideas were applied as far back as \cite{Butcher69} to general Runge-Kutta methods without restriction to Hamiltonian systems.

In particular, we take the same problem of a single planet migrating inward from Sec.\:\ref{secdissipation}, except we shorten the semi-major axis damping timescale $\tau_a$ to 100 (initial) orbital periods, and integrate for one damping timescale.

\begin{figure}
 \centering \resizebox{\columnwidth}{!}{\includegraphics{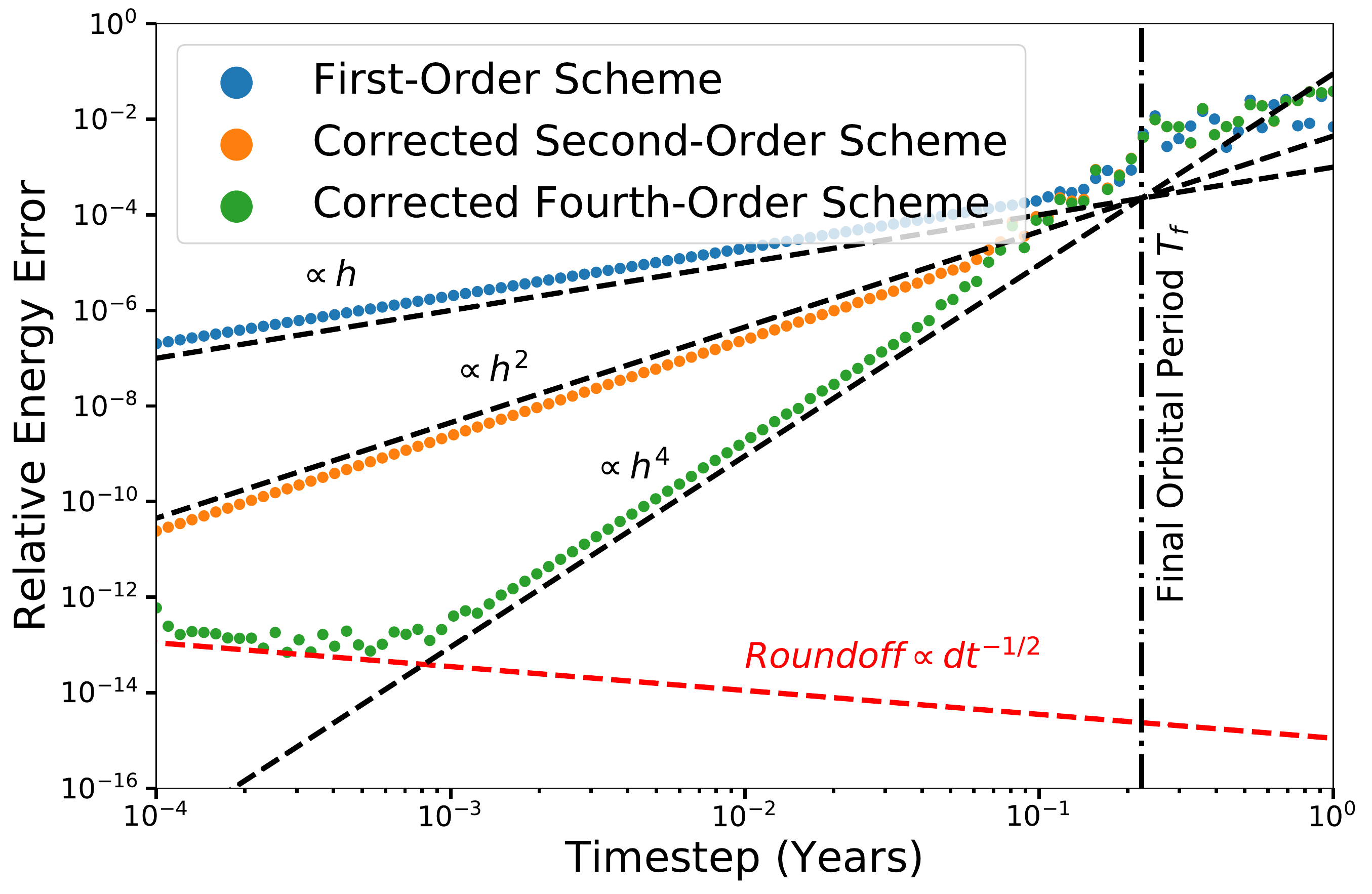}}
 \caption{ Relative energy error in integrations analogous to Fig.\:\ref{figdamp} of a migrating planet, using `symplectic correctors'. Despite the dissipative forces, `symplectic correctors' fix the leading scheme errors to provide the expected scaling. As discussed in the text, `symplectic correctors' are really splitting correctors and provide a way to suppress energy errors by orders of magnitude at fixed computational cost. The fourth-order scheme reaches roundoff errors at small timesteps. 
\label{dampingcorrectors}}
\end{figure}

We consider three integration schemes.
First we apply the first order scheme $\mathcal{K}(h) \circ \mathcal{D}(h)$. From Secs.\:\ref{secEerr} and \ref{secdissipation}, this will yield errors $\sim \epsilon_D (h/T) e$.
We then run a separate integration using first-order correctors (Eq.\:\ref{1stordercorr}), whose error should scale as $\epsilon_D (h/T)^2 e$, and an integration with third-order correctors $\mc{C}_3$ (Eq.\:\ref{itercorr}), which should scale as $\epsilon_D (h/T)^4 e$ for small $h$. 
The results are shown in Fig.\:\ref{dampingcorrectors}, with all curves following the above estimates (black dashed lines).

Several features are apparent from Fig.\:\ref{dampingcorrectors}. 
First, when thought of as a `symplectic' corrector, it might seem surprising that symplectic correctors applied at the beginning and end of the integration (when the energy has dissipated by a factor of 3 in between) would give a more accurate result. 
But as argued above, in our case there is nothing specifically symplectic about them, they are simply weak splitting correctors. 
Second, while all the timesteps considered are small compared to the initial orbital period of 1 year, it is the final orbital period (vertical dashed line) that matters. 
We see that for timesteps $\gtrsim 10\%$ of the {\it final} orbital period, the error rises dramatically, a consequence of higher order terms in the BCH series becoming important.

It might seem counterintuitive that symplectic correctors can improve integrations involving dissipation. 
In particular, one might worry that any steps backwards in time in $\mathcal{C}(h)$ would not cancel with corresponding steps forward in time $\mathcal{C^{-1}(h)}$ when irreversible processes are involved.
For example, a step forward with friction is not undone by a step backward in time---the second step would decrease the energy further.
But this is {\it not} what is meant by, e.g., $\mathcal{A}(-h)$, i.e., it is not a step backward in time in the true sense where all the velocities are also flipped; rather, it merely means running the independent time variable run backwards, keeping everything else the same.
Under this definition, a backwards step with frictional forces {\it boosts} the velocities, undoing a step forward in time, and removing the contradiction.
In the terminology of \cite{Hairer2006}, the scheme is time-symmetric, but not reversible.
\cite{Hernandez18} also find cases where time-symmetric, nonreversible schemes do not undergo energy drifts (see their Fig.\:8).

The fact that `symplectic correctors' can be derived directly from the  BCH formula implies that they must be more general, given that the BCH  formula itself makes no distinction between conservative or dissipative operators.
Finally, the fact that the green curve in Fig.\:\ref{dampingcorrectors} reaches the theoretical round-off limit at $\sim 10^{-14}$ demonstrates that `symplectic correctors' do improve integrations with dissipation to high accuracy.

This points out a limitation of splitting methods for dissipative systems. 
Splitting methods require a fixed timestep (though see, e.g., \citealt{Preto99}), and it must be shorter than the fastest timescale in the problem throughout the integration. 
If there is substantial dissipation, the required timestep might be so small that a high-order adaptive scheme like \ias is more efficient. 
We note that this concern also applies to the conservative N-body problem. 
For example, if Kozai cycles or other effects move planets onto very eccentric paths at late times, the timestep for the whole integration has to be chosen to match the shortest pericentre passage timescale that occurs in the integration\footnote{There are ways to address this using time regularization.}.

\section{Imperfect Operators} \label{velforces}

The above discussion assumes that an analytical form for each step can be found, so that the only errors come from the splitting scheme itself, e.g., Eq.\:\ref{P2nderr}.
However, for many perturbations, analytical solutions are not known, which naturally leads to the question of how errors from numerically approximating the evolution across each step interact with the splitting errors discussed above.

As mentioned above, this is not an issue for position-dependent forces, since they can be trivially incorporated into the interaction step, Eq.\:\ref{diffeqWH}.
In that case Eq.~\ref{vupdate} remains the exact analytical solution, using the accumulated accelerations from all the position-dependent effects. 

However, many important astrophysical effects like the migration forces considered in the previous section, post-Newtonian corrections, tides, etc., are velocity-dependent.
Previous authors have proposed incorporating such velocity-dependent perturbations in either the Kepler step \citep{Malhotra94}, in the interaction step \citep{Cordeiro96}, or as a separate step \citep{Touma94}.
The methods achieve comparable accuracy \citep{Cordeiro96}, so we focus on the methods of \cite{Touma94} and \cite{Cordeiro96}, which fit directly into the framework discussed above.

In this case, the velocity-dependent accelerations vary across the perturbation step as the velocities change, so Eq.\:\ref{vupdate} is no longer exact; it is now merely an Euler step approximation, accurate only to first order in $h$.
In this section we consider the errors that such an Euler step introduces, and compare it to higher-order numerical approximations to the evolution across the perturbation step in this general case where the propagator cannot be found analytically.

\subsection{Hamiltonian Velocity-Dependent Forces} \label{hamvel}

We first consider Hamiltonian velocity-dependent effects, which we will find are qualitatively different from ones involving dissipation.
We note that while we will find such perturbations cause severe numerical problems for these previously proposed methods, those authors were interested in dissipative perturbations, which we consider in Sec.\:\ref{secdissipation}.

A good test case is the velocity-dependent first-order post-Newtonian correction for general relativity \citep{Anderson75}, where, given the dominant central mass in planetary applications, we ignore second-order corrections of order the planet-star mass ratio. The equations of motion can be derived from a Hamiltonian (see Appendix \ref{GRappendix}), whose conservation we can use to track the numerical accuracy.

In all integrations we adopt the first-order splitting 
\begin{equation}
\mathcal{K}(h) \circ \mathcal{GR}(h), \label{GRstep}
\end{equation} 
where the corresponding differential equations are
\begin{align}
\hat{K}{\bf z}&: {\bf \dot{r}_i} = {\bf v_i}, \: &&{\bf \dot{v}_i} = {\bf a_i^{Kep}}({\bf r}) \nonumber \\
\widehat{GR}{\bf z}&: {\bf \dot{r}_i} = 0, \: &&{\bf \dot{v}_i} = \bf a_i^{GR}({\bf r, \bf v}) . \label{GRhat}
\end{align}
with the ${\bf a_i^{GR}}$ given in Eq.\:\ref{GRacc}.

We note that it is possible to find a splitting for this post-Newtonian Hamiltonian for which the evolution under each operator {\it can} be solved analytically \citep{Saha94}.
We nevertheless choose to use the above splitting both to explore the effects of imperfect approximations across the perturbation step, and with a view toward making the \rebx library a general-purpose tool for integration.
Such Hamiltonian velocity-dependent perturbations change the relationship between the particles' physical velocities and their momenta through Hamilton's equations, so this relationship varies depending on what forces are added to an integration. 
To minimize the necessary logic and possible pitfalls, we choose in \rebx to always use the splitting \ref{GRhat}, which can be applied to both conservative and dissipative velocity-dependent forces.
This is the same setup as in both \cite{Touma94} and \cite{Cordeiro96}; we now compare their choice of integrating across the $\mc{GR}$ step using a first-order Euler approximation (Eq.\:\ref{vupdate}) to using higher-order methods.

We first note that in this case of Hamiltonian velocity-dependent forces, \cite{Mikkola98} proposes using a low-order symplectic scheme like the implicit midpoint method to salvage the symplecticity of the scheme.
However, even a symplectic scheme will not be symplectic when applied the non-canonical position and velocity coordinates of Eq.\:\ref{GRhat}, like all the above authors consider\footnote{As a simple analogy, consider a canonical (and thus symplectic) transformation that rotates 2D Cartesian coordinates. If we instead consider a transformation that first converts to non-canonical polar coordinates, naively applies the same symplectic rotation matrix to the polar coordinates, and then converts back to Cartesian coordinates, the sequence of transformations will not be canonical.}.
Therefore, one might not actually expect any advantage from using a symplectic scheme across the perturbation step. 

We test this empirically on the K2-137 system \citep{Smith17}, for which post-Newtonian effects are important.
It consists of a 0.89 Earth-radii ($R_\oplus$) planet on a 4.3 hour ($a_1 \approx 0.01$ AU) orbit around a $0.46$ solar-mass primary.
We assign the planet the mass of the Earth.
While in reality tides should have circularized the orbit, we inflate the initial orbital eccentricity to 0.01, in order to see the apsidal precession induced by the relativistic effects.
The post-Newtonian perturbation strength is $\epsilon_{GR} = 3GM/ac^2 \sim 3 \times 10^{-6}$ \citep[e.g.,][]{Nobili86}.

We compare integrating across the GR perturbation step with the second-order, symplectic implicit midpoint method advocated by \cite{Mikkola98} to three non-symplectic schemes: first-order Euler \citep{Touma94, Cordeiro96}, second-order Ralston's Runge Kutta (RK2), and fourth-order Runge Kutta (RK4).
In all cases the timestep is $\approx 8.1\%$ of the innermost planet's orbital period.
This implies a relative error of $\sim \epsilon_{GR} (h/T) e \sim 2 \times 10^{-9}$ (black arrow).

\begin{figure}
 \centering \resizebox{\columnwidth}{!}{\includegraphics{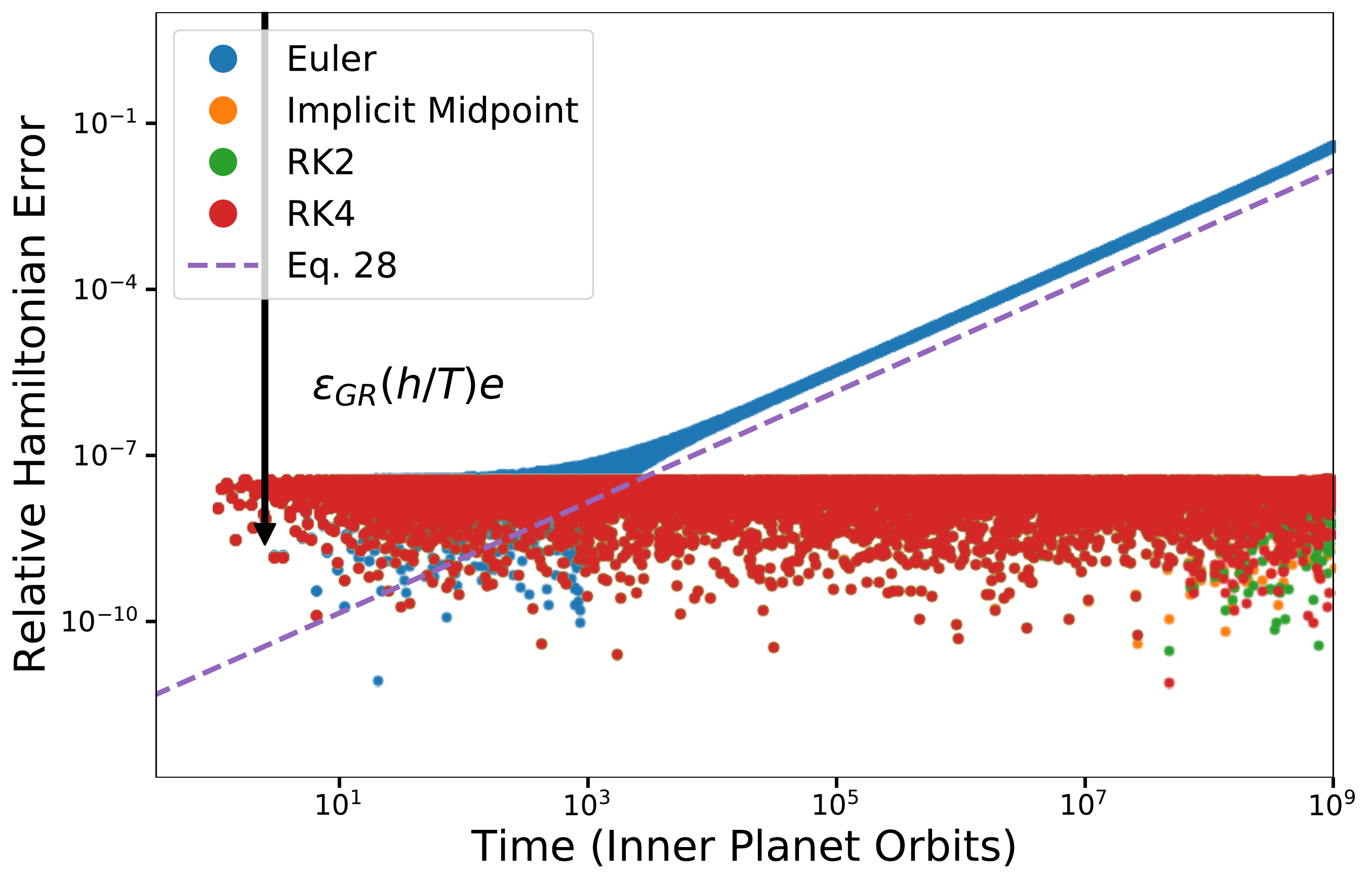}}
 \caption{Integration of the ultrashort period planet K2-137b under relativistic effects with an inflated orbital eccentricity  of 0.01. Naive incorporation of the post-Newtonian accelerations into the Wisdom-Holman scheme (Euler method, blue) leads to a secular error growth, and the innermost planet escapes the system within 6 billion orbits or less than 3 million years. Integrating across the perturbation step using higher order methods, whether they are symplectic (implicit midpoint) or not (RK2, RK4), all give oscillatory errors that remain constant over the length of the integration.
\label{GRplot}}
\end{figure}

In the RK2, RK4 and implicit midpoint integrations in Fig.\:\ref{GRplot}, the truncation errors remain much  smaller than the splitting scheme errors, and the simulations are visually indistinguishable, showing flat, oscillatory errors.
As argued above, applying a symplectic algorithm like the implicit midpoint method to non-canonical variables does not yield a symplectic scheme, so none of these integrations are exactly symplectic. 
Nevertheless, by ensuring that the truncation errors across the perturbation step are much smaller than the splitting scheme errors, we only see the oscillatory symplectic behavior.

Of course, the most salient feature in Fig.\:\ref{GRplot} is that the error in the Euler method of \cite{Touma94} and \cite{Cordeiro96} (blue) error grows secularly, and in fact the planet is spuriously ejected after approximately 5.8 billion orbits, or in under 3 million years.
This behaviour too can be analytically estimated.

\subsection{Truncation Errors} \label{truncerrors}

For general perturbations that can not be integrated analytically, the local truncation error across the perturbation timestep must be incorporated into the error budget.

The local truncation error $\Delta z_{LT}^{\rm{err}}$ across a single timestep using a one-step, order-$n$ method comes from the first neglected term at order $n+1$ in a Taylor series approximation to the solution $z = f(t)$ (we focus on a single scalar coordinate $z$ for simplicity),
\begin{equation}
\Delta z_{LT}^{\rm{err}} = \frac{f^{n+1}(\xi)}{(n+1)!}h^{n+1},
\end{equation}
where $f^{n+1}$ is the $n+1$th derivative of the true solution across the perturbation step, and $\xi$ is an unspecified time in the range $[t, t+h]$. 
If we assume a solution $f(t) = z_0 \exp(i 2\pi t/T_{pert})$, with the characteristic perturbation timescale $T_{pert}$ a factor of $\epsilon$ longer than the orbital timescale $T$, we have
\begin{equation} \label{LTE}
\frac{\Delta z_{LT}^{\rm{err}}}{z} \sim \frac{(2\pi)^{n+1}\epsilon^{n+1}}{(n+1)!}\Bigg(\frac{h}{T}\Bigg)^{n+1}.  
\end{equation}

This is the error incurred every timestep. 
We can obtain the worst-case relative global truncation error after an integration time $t$ by assuming the local errors add coherently, and multiplying Eq.\:\ref{LTE} by the number of steps $t/h$,
\begin{equation} \label{globtrunc}
\frac{\Delta z_{LT}^{\rm{err}}}{z} \sim \frac{(2\pi)^{n+1}\epsilon^{n+1}}{(n+1)!} \Bigg(\frac{h}{T}\Bigg)^n \Bigg(\frac{t}{T}\Bigg)
\end{equation}

We overplot this estimate for the Euler method of \cite{Touma94} and \cite{Cordeiro96} ($n=1$) in Fig.\:\ref{GRplot} and see it gives a good match to the numerical behaviour. The truncation errors for the second-order RK2 method would be lower by a factor of $\epsilon_{GR} (h/T) \sim 10^{-7}$ and thus never become visible.

We might, however, expect different behavior under weak dissipation. 
To reach Eq.\:\ref{globtrunc} we assumed the worst-case scenario that the one-step errors added coherently.
At a more detailed level, one must consider how such errors are propagated by the dynamical flow itself.
Consider a set of trajectories in phase space within a differential error volume around the true trajectory.
By the divergence theorem, the fractional growth of this volume element across a timestep is given by the product of $h$ and the divergence of the vector field of trajectories $\nabla \cdot {\bf \dot{z}}$.
In a Hamiltonian system like above, this divergence vanishes through Hamilton's equations (this is Liouville's theorem), so, at best, the dynamics are neutral for the accumulation of errors\footnote{In reality, an initial parcel of trajectories within some error volume would get sheared out at fixed volume}.

By contrast, under weak dissipation, the divergence of the flow is $\mathcal{O}(\epsilon)$ and negative, so errors contract by $\mathcal{O}(\epsilon h)$ each timestep as the dynamics brings nearby trajectories together. 
One might therefore expect that integrating across the perturbation step with even a first-order Euler method would be good enough, given that one adds one-step errors $\sim \mathcal{O}(\epsilon^2 h^2)$ (Eq.\:\ref{LTE}) slower than they are damped by the dynamics.
We test this empirically in the following section.

\subsection{Dissipation} \label{truncdissipation}
Taking a somewhat realistic damping problem, we initialize two Jupiter-mass planets in a 3:2 mean-motion resonance around a Sun-like star, with the innermost planet initially at 0.90 AU using \celmech\footnote{\celmech is a publicly available code for semi-analytic manipulations of the classical disturbing function in celestial mechanics. Among other things, it can perform transformations between orbital elements and resonant variables: \url{https://github.com/shadden/celmech}}.
We then apply an exponential eccentricity damping force prescription \citep{Papa00} to both planets with an e-folding decay timescale of $\tau_e \approx 1000$ initial inner-planet orbits using the {\tt modify\_orbits\_forces} implementation in \rebx. The outer planet is then additionally acted on by an inward migration force (Eq.\:\ref{Fdrag}) with a timescale 100 times longer $\tau_a \approx 0.1$ Myr. The eccentricities grow as the system evolves deeper into resonance until the migration forces pushing the system into resonance are balanced against the dissipative divergence of the orbits caused by the eccentricity damping \citep[e.g.,][]{Batygin12, Lithwick12, Goldreich14}, reaching equilibrium eccentricities of 0.022 and 0.025 for the inner and outer planet, respectively.

\begin{figure}
 \centering \resizebox{\columnwidth}{!}{\includegraphics{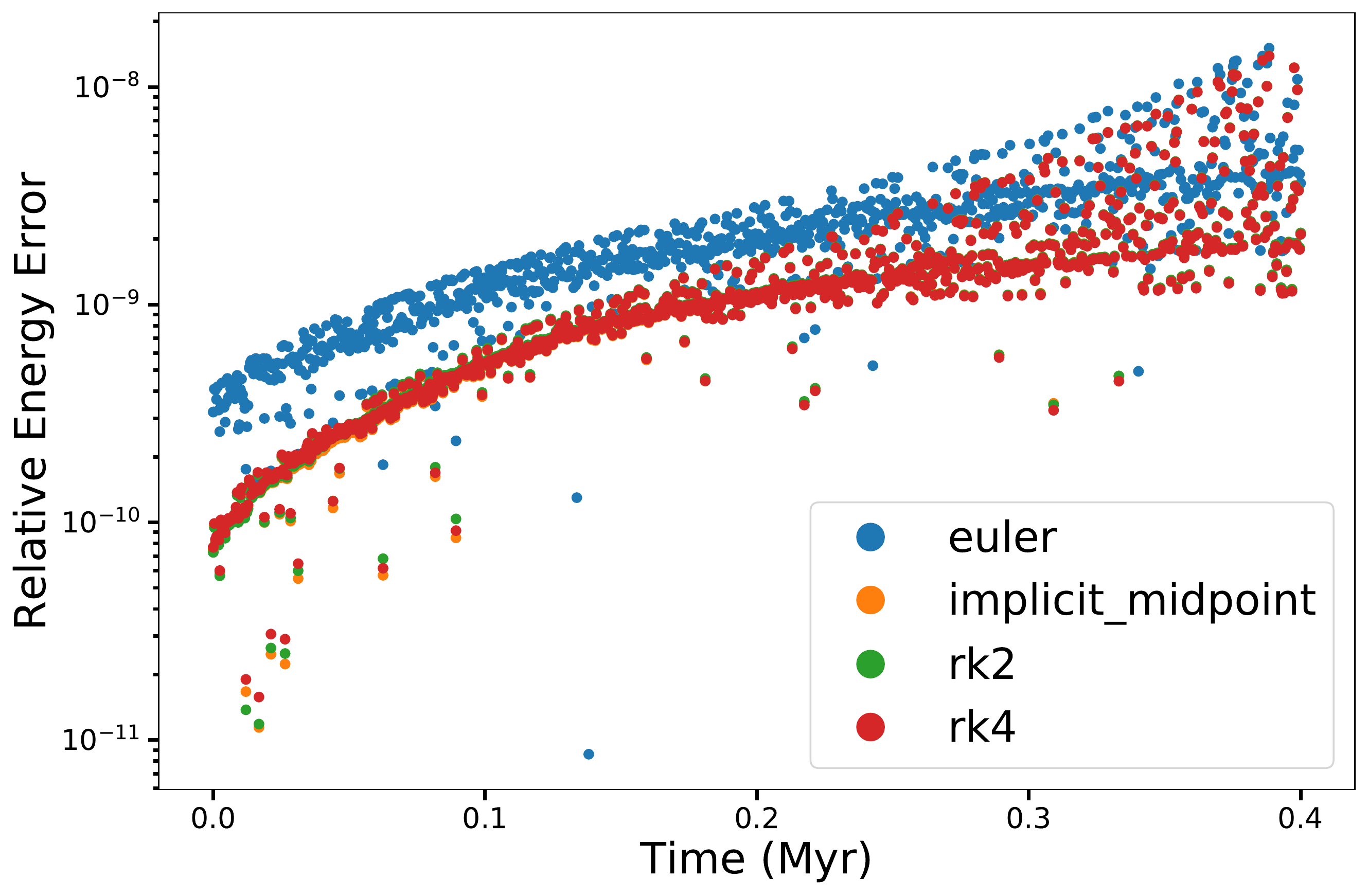}}
 \caption{ Integration of two Jupiter-mass planets in a 3:2 mean-motion resonance equilibrium between migration forces pushing the system deeper into resonance (and both planets inward), and eccentricity damping pushing the system out of resonance. Inner planet starts with an orbital  period of $\approx 1$ year, shrinking to $\approx 0.04$ years by the end of the simulation. We integrate across the velocity-dependent damping step using four different methods, and all give comparable results. Contrast with Fig.\:\ref{GRplot}, where the Euler method fails. See text for discussion.
\label{figdamp}}
\end{figure}

In Fig.\:\ref{figdamp} we then plot the energy errors during the subsequent integration, where the eccentricities remain in a quasi-steady state as the two plants migrate inward together, on a timescale approximately two times longer than the $\tau_a$ estimate above (two equal mass planets have to be moved). We use the second-order scheme\footnote{We provide a simple argument why one can always use such a time-symmetric splitting with multiple operators to create a second-order scheme. The timestep coefficients for each operator must sum to unity in order to match the modified differential equations to the true ones at zeroth order, and by writing an explicitly time-symmetric scheme, we ensure that the additional terms in the modified differential equations cannot depend on the sign of $h$ (e.g., Eq.\:\ref{P2nderr}). The scheme is therefore second-order by construction.}
\begin{equation}
\mathcal{K}(h/2) \circ \mathcal{D}(h/2) \circ \mathcal{I}(h) \circ \mathcal{D}(h/2) \circ \mathcal{K}(h/2), \label{dampingscheme}
\end{equation}
and integrate across each damping step using the four methods of Sec.\:\ref{hamvel}. We integrate for $4\times10^5$ initial orbits, which translates to approximately 4000 eccentricity-damping timescales, or 2 semimajor-axis damping timescales. We adopt a small fixed timestep of $10^{-4}$ of the innermost planet's initial orbital period to accommodate the shrinking orbital periods, which decrease by a factor of $\approx 25$ over the integration. As above, we compare the energies to the energies in a `perfect' integration with \ias.

In this case the scheme errors are dominated by the interplanetary, resonant interactions.
As in Sec.\:\ref{secdissipation}, we see in Fig.\:\ref{figdamp} that the energy errors grow due to the shrinking orbital periods.
Like in Fig.\:\ref{GRplot}, the RK2, RK4 and implicit midpoint integrations are visually indistinguishable. 
By contrast, while the integrations using the first-order Euler method across the damping step have higher errors, it is not a runaway effect as in the conservative case in Fig.\:\ref{GRplot}.
As discussed in Sec.\:\ref{truncerrors}, this is because of the dissipative dynamics continually damping any accumulated energy errors.

In summary, the crudest, most computationally efficient Euler step across a velocity-dependent force step should give reliable results for dissipative forces, as advocated by \cite{Touma94} and \cite{Cordeiro96}.
However, such a scheme would yield qualitatively wrong results in long-term integrations with conservative velocity-dependent forces (Sec.\:\ref{hamvel}).
This can straightforwardly be solved by integrating across the perturbation step with a higher method; for high-accuracy explicit splitting schemes that go beyond this to correct even such higher-order integrations across the perturbation step, see \cite{Blanes13}.

\section{REBOUNDx} \label{sec:rebx}
In the previous sections we covered several technical aspects of how additional forces interact with splitting integration schemes like \whfast.
However, many readers are likely interested in more practical questions of how to quickly get something working.
We therefore implement the integration schemes and ideas discussed above in a new library, \rebx, which can be found at \url{https://github.com/dtamayo/reboundx}. 

\rebx is library that provides tools and routines to accurately and self-consistently incorporate many different astrophysical effects to $N$-body simulations.
It seamlessly interfaces with the \reb N-body integration package \citep{REBOUND}. 
Like \reb, \rebx is written in \clang, and provides a \python wrapper to interface with other libraries. 
Also, like \reb, the \rebx source code is machine independent.
We implement a binary format to save \rebx configurations that interfaces with the {\tt SimulationArchive} class in \reb, making it possible to share and reproduce results bit by bit. 

We have already implemented several common forces in \rebx, some of which we used as examples in the discussion above.
We hope to increase the number of supported astrophysical effects over time.
\rebx is an open-source project and we welcome the community to contribute new effects.
We provide many tutorials and examples in the form of Jupyter notebooks that illustrate the library's usage and how one can add new effects with minimal effort.
We also share the scripts used to generate all the plots in this paper in a separate repository at \url{https://github.com/dtamayo/reboundxpaper}.

We now give an overview of the main concepts, structures and logic in the \rebx package.


\subsection{Which Integrator To Use?}
In a scenario where all the forces (including interplanetary forces) are always small compared to the gravitational forces from the central body, we recommended using Wisdom-Holman Kepler splitting schemes.
The implementation of this method in \reb is \whfast \citep{WHFAST}.
This splitting will typically be fastest at reasonable levels of accuracy, and it provides strong conservation properties in long-term integrations.

Higher levels of accuracy are supported in cases where the forces can all be derived from position-dependent potentials by setting {\tt sim.ri\_whfast.corrector} $= N$, where $N$ is the order of the corrector (Sec.\:\ref{sympcorr}).
In velocity-dependent cases, correctors can be implemented manually (the notebook {\tt fig4.ipynb} in this paper's repository provides an example).
For even higher levels of accuracy (without significant computational cost), one can use the WHCKL method \citep{Rein2019b}, which is an implementation of the kernel method of \cite{Wisdom96} \citep[see also][]{Wisdom18} that also removes errors at $\mathcal{O}(\epsilon^2)$.

In cases with close encounters between planets, the motion will no longer be nearly Keplerian, and Kepler splittings will yield large errors.
In cases where such close encounters are rare, hybrid integrators are powerful tools \citep{Duncan98, Chambers99, MERC}.
In \reb, the \mer integrator retains efficiency by using a Kepler splitting when possible, and switches to an adaptive, high-accuracy integration with \ias for particles undergoing close encounters \citep{MERC}.
\mer supports additional forces, but only switches to \ias during close gravitational encounters.
In the current implementation, any additional forces must remain small throughout the integration when \mer is used.

When the above conditions are not met, the safest option (and the default) in \reb is to use the \ias integrator \citep{IAS15}.
Because it is an adaptive scheme, \ias can handle arbitrary forces and still yield solutions that are accurate close to the machine precision.
\ias will typically be slower in cases where \whfast and \mer can be used.
However, \ias will always be more accurate and is applicable to a wider variety of problems\footnote{
By using adaptive timesteps, \ias can even be faster in cases, e.g., very eccentric orbits, where one has to choose a small timestep with \whfast or \mer in order to resolve a short timescale in the problem (e.g., pericenter passage).}.

\subsection{\rebx Structures}
The two principal ways to incorporate additional effects into an $N$-body simulation are through forces and operators.

\subsubsection{Forces}
One way to implement an additional effect is to write it explicitly as a force.
This force then contributes to the accelerations of all particles, in addition to the standard Newtonian gravity.

When the \ias integrator is used, the integrator automatically adapts the timestep such that integrations will be accurate close to machine-precision.
If one of the various splitting integrators available within \reb (for example \wh, \mer, and \lf) is used, then the additional forces are accumulated and applied in the integrator's interaction (or kick) step using Eq.\:\ref{vupdate}.
As discussed above, this is appropriate for position-dependent or dissipative forces (Secs.\:\ref{posforces} and \ref{truncdissipation}).
For conservative, velocity-dependent forces, one should add a separate operator (Sec.\:\ref{rebxoperator}).

For the hybrid symplectic \mer integrator \citep{MERC}, only forces derivable from position-dependent potentials are supported.

\subsubsection{Operators} \label{rebxoperator}
Rather than calculating accelerations which are then integrated in the interaction or kick step of the integrator, operator steps yield solutions. 
Operators update the positions and velocities (or even masses or other particle parameters) to the value they should have at the end of a specified timestep. 

Operators (e.g., $\mc{D}$) together with a timestep (e.g., $h/4$) form a step (e.g., $\mc{D}(h/4)$).
The same operator can thus be used in multiple steps with different timesteps to construct higher-order splitting schemes (e.g., Eq.\:\ref{dampingscheme}). 

For cases where the solution for the positions and velocities at the end of the step is not known, or difficult to calculate analytically, the \rebx object {\tt integrate\_force} operator can take a force, and integrate it across the timestep using any of the four integrators discussed in Sec.\:\ref{velforces}. 
The default is RK4.
This is the recommended method for conservative, velocity-dependent forces (Sec.\:\ref{hamvel}).

\subsection{Ordering}
All forces and operator steps are stored in linked lists such that they are executed in the reverse order they were added. 
Thus, to get a scheme consisting of the steps $\mathcal{K}(h) \circ \mathcal{D}(h)$, the operator $\mathcal K$ should be added before $\mathcal D$, i.e.  left to the right, since the last step acts first. 

\subsection{Timestep logic}
\rebx provides a simple API for adding effects into N-body integrations from the list of implemented options, demonstrated in the numerous examples in the documentation.
However, it also provides lower-level functionality for customized tasks. 
The three main function pointers in a \reb {\tt Simulation} object that \rebx sets and calls behind the scenes every timestep are
\begin{itemize}[leftmargin=*]
    \item {\tt additional\_forces}. This iterates through the linked list of attached forces to update particle accelerations.
    \item {\tt pre\_timestep\_modifications}. This iterates through the linked list of attached operator steps to execute before the main \reb step.
    \item {\tt post\_timestep\_modifications} This iterates through the linked list of operator steps to execute after the main \reb step.
\end{itemize}
The user populates these lists by either adding forces and operators already implemented in \rebx, or through user-defined functions in either \python or \clang.
By default, adding an operator to an integration adds half steps before and after the main \reb step to make a second-order scheme.

In Fig.~\ref{fig:schema}, we schematically outline a full \reb timestep using the \whfast integrator.
Different integrators switch out the {\tt reb\_integrator\_part1} and {\tt reb\_integrator\_part2} steps, but in all cases they rely on the gravitational and additional forces calculated in between them.

\tikzstyle{normal} = [rectangle, draw, rounded corners, text width = 12em,  align=center]
\tikzstyle{rebxfunction} = [rectangle, draw,  rounded corners, fill=black!15,  text width = 12em, align=center]
\begin{figure}
    \centering
\begin{tikzpicture}
\node (pre) [normal] { are pre-timestep \\modifications present?};
\node (dopre1) [normal, below right = 0cm and 1cm of pre] { convert coordinates to inertial frame};
\node (dopre2) [rebxfunction, below = 5mm of dopre1] { apply pre-timestep modifications};
    \node (part1) [normal, below = 2.7cm of pre] { WHFast Kepler step\\(half timestep)};
\node (grav) [normal, below = 9mm of part1] { calculate gravitational accelerations};
\node (af) [normal, below of=grav] { are additional\\ forces present?};
\node (doaf1) [normal, below right = 0cm and 1cm of af] { convert coordinates to inertial frame};
\node (doaf2) [rebxfunction, below = 5mm of doaf1] { calculate \\additional forces};
    \node (part2a) [normal, below = 2.9cm of af] { WHFast interaction step\\ (full timestep)};
    \node (part2b) [normal, below = 4mm of part2a] { WHFast Kepler step \\(half timestep)};
\node (post) [normal, below = 4mm of part2b] { are post-timestep modifications present?};
\node (dopost1) [normal, below right = 0cm and 1cm of post] { convert coordinates to inertial frame};
\node (dopost2) [rebxfunction, below = 5mm of dopost1 ] { apply post-timestep modifications};
\node (col) [normal, below = 2cm of post] {search for collisions\\ and ejections};
\draw [arrow] (pre) -- node[anchor=south] {Yes} (dopre1);
\draw [arrow] (pre) -- node[anchor=east] {No} (part1);
\draw [arrow] (dopre1) -- (dopre2);
\draw [arrow] (dopre2) -- (part1);
\draw [arrow] (part1) -- (grav);
\draw [arrow] (grav) -- (af);
\draw [arrow] (af) -- node[anchor=south] {Yes} (doaf1);
\draw [arrow] (af) -- node[anchor=east] {No} (part2a);
\draw [arrow] (doaf1) -- (doaf2);
\draw [arrow] (doaf2) -- (part2a);
\draw [arrow] (part2a) -- (part2b);
\draw [arrow] (part2b) -- (post);
\draw [arrow] (post) -- node[anchor=south] {Yes} (dopost1);
\draw [arrow] (dopost1) -- (dopost2);
\draw [arrow] (dopost2) -- (col);
\draw [arrow] (post) -- node[anchor=east] {No} (col);
\node (part1box) [inner sep=1em, draw,thick,dotted,fit=(part1)] {};
    \node [right = 2mm of part1box.east, anchor = west] {{\tt reb\_integrator\_part1}};
\node (part2box) [inner sep=1em, draw,thick,dotted,fit=(part2a) (part2b)] {};
\node [right = 2mm of part2box.east, anchor = west] {{\tt reb\_integrator\_part2}};
\node (updateaccbox) [inner sep=1em, draw,thick,dotted,fit=(grav) (doaf1) (doaf2)] {};
\node [above right = -3mm and -1mm of updateaccbox.north east, anchor = east] {{\tt reb\_update\_acceleration}};
\end{tikzpicture}
 \caption{Schematic outline of a full \reb step when using the \whfast integrator.
    Functions which are part of \rebx and called via function pointers are shaded.
\label{fig:schema}}
\end{figure}
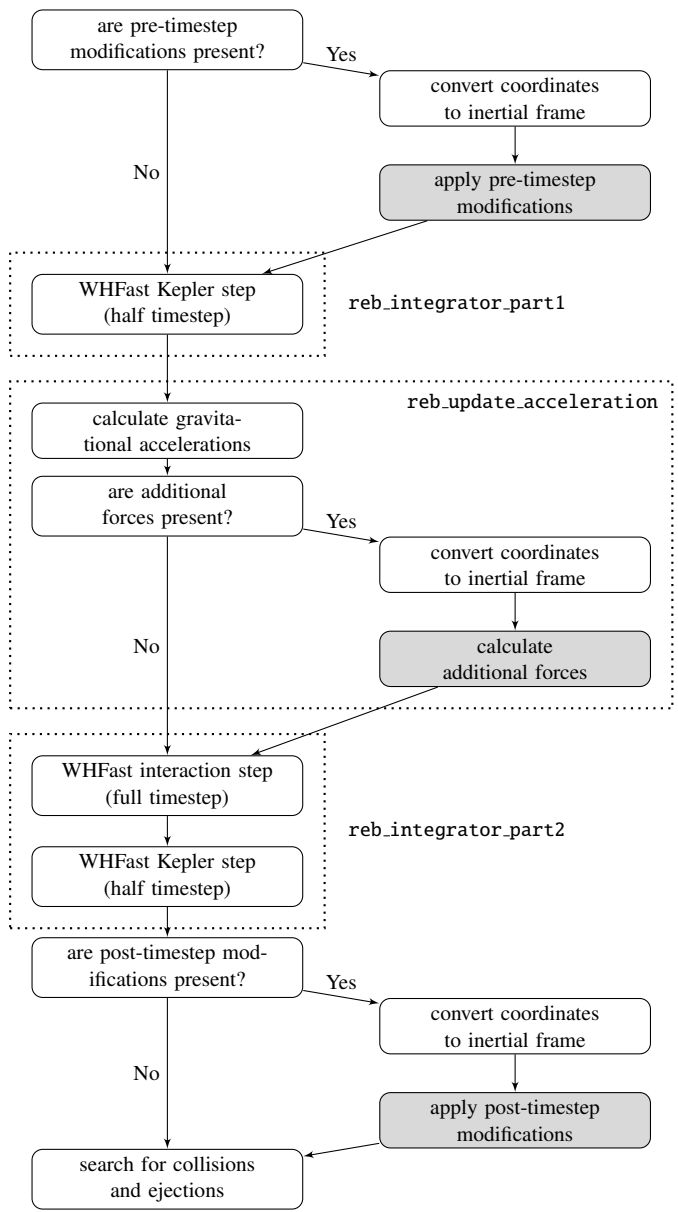

\subsection{Coordinate systems}
In contrast to many other N-body packages, \reb allows the user to set up the simulation in an arbitrary inertial frame.
The coordinates are returned to the user in the same inertial frame.

The various splitting integrators in \reb use a number of different coordinates internally (e.g. Jacobi coordinates).
In order to minimize logic and pitfalls, all \rebx functions are always called after converting the particles' positions and velocities back to Cartesian coordinates in the inertial frame.
Force and operator implementations thus do not need to worry about indirect or non-inertial forces.
This also implies that it is typically desirable to ensure that the net applied forces vanish.
This is discussed in detail in Appendix \ref{COM}.

\subsection{Parameters}
Additional effects in an $N$-body simulation typically come with several associated parameters (e.g., a migration timescale for a damping force). 
\rebx provides an interface which allows a user to attach an arbitrary numbers of parameters to forces, operators, and particles.
The implementation is such that parameters remain tied to their parent structures even as objects get removed or reshuffled in the simulation.

\subsection{Implemented effects}
A current list of implemented effects can be found in the documentation.
At the time of this writing, the following effects are available:
\begin{itemize}[leftmargin=*]
    \item Migration forces \citep{Papa00}
    \item Eccentricity and inclination damping forces \citep{Papa00}
    \item General relativistic 1 PN corrections \citep{Newhall83}
    \item General relativistic 1 PN corrections for a dominant central mass \citep{Anderson75}
    \item General relativistic 1 PN corrections with a simplified potential \citep{Nobili86}
    \item Radiation Forces \citep{Burns79}
    \item Mass loss
    \item Precession from conservative equilibrium tides \citep{Hut81}
    \item General central forces
    \item Higher order gravitational moments, J2 and J4 
\end{itemize}

\section{Conclusion} \label{conclusion}

Throughout this paper, we have developed the error behaviour of splitting schemes in a general, non-Hamiltonian framework. 
This approach clarifies how several properties typically ascribed to symplectic integrators also carry over to weakly dissipative systems.

Many astrophysically relevant effects have strong symmetry properties. 
We showed explicitly that the remarkable degeneracy of the Kepler problem makes schemes that split the evolution into Kepler steps and perturbation steps exploit these symmetries to often strongly suppress energy errors in the integration, as in the cases of radial or drag forces applied to nearly circular orbits (Sec.\:\ref{geom}).
We also showed that so-called `symplectic correctors', which reduce energy errors by orders of magnitude at fixed computational cost \citep{Wisdom96}, apply equally well to weakly dissipative systems and can thus be more generally thought of as `weak splitting correctors' (Sec.\:\ref{seccorrectors}).

We also explored the general case where individual operator steps can't be solved analytically, and one is forced to integrate across the perturbation step.
We showed that the previously advocated approaches of incorporating additional forces into the Wisdom-Holman map \citep{Touma94, Malhotra94, Cordeiro96} work well for dissipative effects (Sec.\:\ref{truncdissipation}), but give qualitatively wrong answers for conservative, velocity-dependent forces like post-Newtonian corrections of general relativity (Sec.\:\ref{hamvel}).

Finally, we described \rebx, an open-source \clang library for incorporating additional effects into \reb N-body integrations, together with a convenient \python wrapper.
Users can either choose to add an already implemented astrophysical forces, or easily implement new effects.
We hope that this modular, open-source framework for robustly incorporating new effects into a variety of N-body integration schemes will help the community to run more accurate and realistic simulations.
We encourage others to contribute to this library of astrophysical effects.

\section*{Acknowledgments}
We would like to thank John Chambers for an insightful and constructive review that greatly improved this manuscript. 
We would also like to thank Jack Wisdom and Scott Tremaine for helpful discussions.
Support for this work was provided by NASA through the NASA Hubble Fellowship grant HST-HF2-51423.001-A awarded  by  the  Space  Telescope  Science  Institute,  which  is  operated  by  the  Association  of  Universities  for  Research  in  Astronomy,  Inc.,  for  NASA,  under  contract  NAS5-26555.
This research has been supported by the NSERC Discovery Grant RGPIN-2014-04553 and the Centre for Planetary Sciences at the University of Toronto Scarborough.
This research was made possible by the open-source projects 
\texttt{Jupyter} \citep{jupyter}, \texttt{iPython} \citep{ipython}, 
and \texttt{matplotlib} \citep{matplotlib, matplotlib2}.


\bibliographystyle{mnras}
\bibliography{Bib}

\appendix

\section{Operator theory} \label{backgroundappendix}
While not strictly necessary for the development in the main text, we review here some of the gaps glossed over in Sec.\:\ref{background}.
Using the same notation, we seek a solution to the differential equations
\begin{equation}
\dot{\z} = \L{P}\z,
\end{equation}
where here we define the Lie derivative more carefully.
The Lie derivative with respect to a set of differential equations $\dot{\z} = \hat{F} \z$ acts on functions $g(\z)$ to return their total derivative. Since throughout the paper we treat only cases that are explicitly time-independent, we have
\begin{equation}
\L{F} g(\z) = \sum_i \frac{\partial{g}}{\partial{z_i}} \dot{z}_i, \label{Llie}
\end{equation}
where the $\dot{z}_i$ are given by $\hat{F} \z$.

With this definition, we can Taylor expand the solution around the current time
,
\begin{equation}
\z\Bigg|_{t+h} = \z\Bigg|_t + h\frac{d\z}{dt}\Bigg|_t + \frac{h^2}{2!}\frac{d^2\z}{dt^2}\Bigg|_t + ... \equiv \Bigg[(1 + h\mc{L}_{P} + \frac{h^2}{2!}\mc{L}_{P}^2 + ...)\z\Bigg]_t. \label{taylor}
\end{equation}
As a concrete example, consider the simple one-dimensional differential equation $\dot{z} = z^2$. We have $\mc{L}_{P}z = z^2$, $\mc{L}_{P}^2z = \mc{L}_{P}z^2 = 2z^3$ etc.. Plugging into Eq.\:\ref{taylor} yields the Taylor expansion of the exact solution $z(t) = (1/z(0) - t)^{-1}$.

By analogy to the Taylor expansion for the exponential, this is often written as\footnote{Throughout the paper we assume that the differential equations do not depend explicitly on time. If they did, an integral would be required in the exponential of Eq.\:\ref{hP}.}
\begin{equation}
{\bf z}(t+h) \equiv e^{h\mc{L}_{P}}{\bf z}(t) \equiv \mathcal{P}(h){\bf z}(t). \label{hP}
\end{equation}
In words, $\mathcal{P}(h)$ is a time evolution operator, i.e., an integrator, that updates the phase space variables across a timestep. 
We then proceed to split the differential equations into two pieces $\dot{\z} = \L{P}\z = (\L{A} + \L{B})\z$ as in the main text.

One subtle notational issue that has led to some errors in the literature is that the BCH formula is typically written as
\begin{equation} \label{expcomp}
e^{h\L{A}}e^{h\L{B}} = e^{h\L{S}}
\end{equation}
with 
\begin{equation} \label{bch}
\L{S} = \L{A} + \L{B} + \frac{h}{2}[\mc{L}_A, \mc{L}_B] + \frac{h^2}{12}[\mc{L}_A - \mc{L}_B, [\mc{L}_A, \mc{L}_B]] + \mathcal{O}(h^3).
\end{equation}
However, $e^{h\L{A}}e^{h\L{B}}$ is {\it not} equivalent to the splitting scheme one might naively implement on a computer $e^{h\L{A}} \circ  e^{h\L{B}}$.

In $e^{h\L{A}}e^{h\L{B}}\z$, $e^{h\L{A}}$ acts on a function $g({\bf z}) = e^{h\L{B}}\z$, and therefore introduces partial derivatives of $g$ through Eq.\:\ref{Llie}.
By contrast, the integrator $e^{h\L{A}} \circ  e^{h\L{B}}$ first evaluates an intermediate ${\bf z}' = e^{h\L{B}}{\bf z}$, and these updated values are used to yield a final answer of $e^{h\L{A}}{\bf z'}$.
The BCH formula applies to the former sense (one can straightforwardly expand the three exponentials in Eq.\:\ref{expcomp} to obtain the first terms in Eq.\:\ref{bch}), but this interpretation is not particularly useful for writing integrators.
Under iterated mappings, the repeated chain rules would quickly become unwieldy.
By contrast, $e^{h\L{A}} \circ  e^{h\L{B}}$ is straightforward to implement on a computer, and fortunately can be simply related to the scheme in the BCH formula to analyze its error properties.
This is sometimes referred to as the Vertauschungssatz \citep{Grobner1967} (see Lemma 5.1 in chapter III of \cite{Hairer2006}); we provide a simple derivation and example here.

Looking back at Eq.\:\ref{taylor}, one could just as well have Taylor expanded any scalar function $f(\z)$, and concluded in Eq.\:\ref{hP} that the time evolution operator $e^{h\L{A}}$ also evolves $f(\z)$ from time $t$ to time $t+h$ (with $f({\bf z}) = {\bf z}$ in Eq.\:\ref{taylor} as a special case). 
Thus, since Eq.\:\ref{expcomp} is of the form  $e^{h\L{A}}f({\bf z})$, with $f({\bf z}) = e^{h\L{B}}{\bf z}$, we must have
\begin{equation}
e^{h\L{A}}f(\z) = f \circ (e^{h\L{A}}\z),
\end{equation}
i.e., evolving $f({\bf z})$ across the timestep (LHS) is equivalent to first evolving ${\bf z}$ and applying the function $f$ to the updated variables (RHS). 
Therefore, 
\begin{equation} 
e^{h\L{A}} \circ e^{h\L{B}} = e^{h\L{B}}e^{h\L{A}}. \label{flip}
\end{equation}
This is the reason for the sign flips at all odd orders in Eq.\:\ref{bchflip}.

As one might expect, Eq.\:\ref{flip} has led to some confusion in notation and sign errors in the literature (e.g., footnote 3 in \citealt{Saha92} and Eq. 10 in \citealt{Hernandez17}).
However, one reason why this is probably not typically noticed or pointed out is that most integrators used in the astrophysics literature are time reversible, and thus will not have any sign flips in their corresponding BCH formulas given that they are invariant to flipping the order of operations (e.g., \citealt{Yoshida93} and the detailed analysis in \citealt{Hernandez17}).

As an explicit example of the above, consider a non-dimensionalized simple harmonic oscillator with position $x$, momentum $v$, and ${\bf z} = \left(x, v\right)$. 
The Hamiltonian is $H = v^2/2 + x^2/2$, and we use a kinetic-potential splitting $H_T = v^2/2$ and $H_V = x^2/2$.
The corresponding differential equations are given by Hamilton's equations,
\begin{eqnarray}
\L{T} \z : \dot{x}_T = \frac{\partial H_T}{\partial v} = v,\:\:\:\:\:\dot{v}_T = -\frac{\partial H_T}{\partial x} = 0, \\
\L{V} \z : \dot{x}_V = \frac{\partial H_V}{\partial v} = 0,\:\:\:\:\:\dot{v}_V = -\frac{\partial H_V}{\partial x} = -x. \\
\end{eqnarray}
The propagators can easily be found by integrating their respective differential equations, or by noting $\L{T}^n\z = \L{V}^n\z = 0$ for $n \geq 2$, so
\begin{eqnarray}
e^{h\L{T}}{\bf z} = (1 + h\L{T}){\bf z} = \begin{pmatrix} x + hv \\ v \end{pmatrix} \\
e^{h\L{V}}{\bf z} = (1 + h\L{V}){\bf z} = \begin{pmatrix} x \\ v - hx \end{pmatrix}.
\end{eqnarray}
We first evaluate the composition used in the $\mc{S}\mc{A}\mc{B}$ integrator  (Eq.\:\ref{1storderKT}), $e^{h\L{T}} \circ e^{h\L{V}}\z$, where $e^{h\L{T}}\z'$ is evaluated at $\z' = e^{h\L{V}}\z$
\begin{eqnarray}
(e^{h\L{T}} \circ e^{h\L{V}})\z = \begin{pmatrix} x + h(v - hx) \\ v - hx \end{pmatrix}. \label{z1update}
\end{eqnarray}
By Eq.\:\ref{flip}, this should be equivalent to 
\begin{eqnarray}
e^{h\L{V}}e^{h\L{T}}{\bf z} = (1 + h\L{V})\begin{pmatrix} x + hv \\ v \end{pmatrix} = \begin{pmatrix} x + hv \\ v \end{pmatrix} + h\begin{pmatrix} \dot{x}_V + h\dot{v}_V \\ \dot{v}_V\end{pmatrix},
\end{eqnarray}
which indeed matches Eq.\:\ref{z1update}.

\section{First Order Post-Newtonian Corrections} \label{GRappendix}

General relativistic corrections to Newtonian dynamics are detailed throughout the literature, but since we use these effects as one of the main tests for the various schemes in this paper, we present the equations and details of our implementation for completeness.

The first order post-Newtonian corrections (1PN) are $\mathcal{O}(v^2/c^2)$, where $v$ is the characteristic orbital velocity and $c$ is the speed of light.
We have implemented various levels of approximation.
In {{\sc \tt gr\_full}\xspace}, we include the full (1 PN) equations of motion, which are implicit and computationally expensive, but would be appropriate for equal-mass bodies.

For planets around a single star, one can typically ignore corrections to the GR perturbation of order the planet-star mass ratio.
This significantly simplifies the equations, but involves velocity-dependent accelerations.
This approximation is implemented in the {{\sc \tt gr}\xspace} effect in \rebx, and is the case discussed in Sec.\:\ref{velforces}.

As a final level of approximation, one is typically interested in the apsidal precession effect from general relativity, in which case it is possible to write down a potential only depending on the particles' position, which, in an orbit-averaged sense gives the correct precession rate \citep{Nobili86}. 
This is implemented in {{\sc \tt gr\_potential}\xspace}, and has the advantage that it can be integrated analytically (Sec.\:\ref{posforces}), and is therefore computationally cheapest.
However, this approximation introduce errors in the phases and instantaneous elements of $\mathcal{O}(v^2/c^2)$.
Typically, {{\sc \tt gr\_potential}\xspace} is sufficient for planetary applications around a single star since it captures the correct secular behavior, but both {{\sc \tt gr\_potential}\xspace} and {{\sc \tt gr}\xspace} are only appropriate for single-star systems. 
For higher multiplicity systems, one must use {{\sc \tt gr\_full}\xspace}.

\cite{Naoz13} provide the full Hamiltonian for a triple system \citep[see also][]{Schafer87}. 
The equations of motion are given by \cite{Newhall83} and \cite{Benitez08}.
These accelerations and Hamiltonian are implemented in {{\sc \tt gr\_full}\xspace}.

In the {{\sc \tt gr}} implementation, we assume a dominant central mass, and ignore additional corrections of order the planet-star mass ratio.
We outline our implementation for this effect in more detail since it differs from previous works, and is the subject of Sec.\:\ref{velforces}.
The full Hamiltonian can be written as a sum of the Newtonian and 1PN Hamiltonians, $\mathcal{H} = \mathcal{H}_{N} + \mathcal{H}_{PN}$, where
\begin{equation}
\mathcal{H}_N = \sum_i \frac{p_i^2}{2m_i} - \frac{1}{2}\sum_{j \neq i} \frac{Gm_i m_j}{r_{ij}}
\end{equation}
and \citep{Saha94},
\begin{equation}
\mathcal{H}_{PN} = \frac{1}{c^2}\sum_{i\neq 0} \Bigg(\frac{\mu^2 m_i}{2r_i^2} - \frac{p_i^4}{8m_i^3} - \frac{3\mu p_i^2}{2m_ir_i}\Bigg)
\end{equation}
Having already ignored terms of $\mathcal{O}(m_i/m_0)$, the difference between barycentric, Jacobi or democratic heliocentric coordinates (and Jacobi or physical masses) in $\mathcal{H}_{PN}$ is negligible.
Interpreting all positions and momenta as Jacobi coordinates considerably simplifies the equations of motion, since the kinetic term in $\mathcal{H}_N$ remains a diagonal sum of the momenta.
Application of Hamilton's equations yields \citep[e.g.,][]{Saha94},
\begin{eqnarray} \label{rdotgr}
\mathbf{\dot{r}_0} &=& \nabla_\mathbf{p_0} \mathcal{H} = \frac{\mathbf{p_0}}{m_0} \equiv \mathbf{\tilde{v}_0}, \nonumber \\
\mathbf{\dot{r}_{i\neq0}} &=& \nabla_\mathbf{p_i} \mathcal{H} = \frac{\mathbf{p_i}}{m_i} + \nabla_\mathbf{p_i} \mathcal{H}_{PN} \equiv \mathbf{\tilde{v}_i} \left(1 + A_i\right),
\end{eqnarray}
where 
\begin{equation}
A_i = -\frac{1}{c^2} \left(\frac{\tilde{\mathrm{v}}_i^2}{2} + \frac{3\mu}{r_i}\right)
\end{equation}
and the $\mathbf{\tilde{v}_i} \equiv \mathbf{p_i}/m_i$ are pseudo-velocities not equal to the physical velocities $\mathbf{\dot{r}_i}$. 
We note that Eq.~\ref{rdotgr} would have additional terms in democratic heliocentric coordinates due to kinetic cross terms in $\mathcal{H}_N$.
We then have
\begin{eqnarray} \label{GRacc}
\mathbf{\ddot{r}_0} &=& \mathbf{\dot{\tilde{v}}_0} = 0, \nonumber \\ 
\mathbf{\ddot{r}_{i \neq 0}} &=& \mathbf{\dot{\tilde{v}}_i} \left(1 + A_i\right) + \dot{A_i}\mathbf{\tilde{v}_i} 
\end{eqnarray}
where
\begin{equation}
\mathbf{\dot{\tilde{v}}_{i \neq 0}} = - \frac{1}{m_i} \nabla_\mathbf{r_i} \mathcal{H} = \mathbf{a_i} + \frac{\mathbf{r_i}}{c^2}\Bigg(\frac{3\mu\tilde{\mathrm{v}}_i^2}{r_i^3} - \frac{\mu^2}{r_i^4}\Bigg),
\end{equation}
$\mathbf{\tilde{a}_i}$ are the net Newtonian accelerations on particle $i$, and 
\begin{equation}
\dot{A_i} = - \frac{1}{c^2}\Bigg(\mathbf{\tilde{v}_i} \cdot \mathbf{\dot{\tilde{v}}_i} - \frac{3\mu}{r_i^3} \left(\mathbf{r_i} \cdot \mathbf{\dot{r}_i}\right)\Bigg).
\end{equation}
We see that the accelerations (Eq.\:\ref{GRacc}) are explicitly velocity-dependent, and must be treated with care in symplectic schemes (Sec.\:\ref{velforces}).
As expected, $\mathbf{\ddot{r}_0} = 0$ and the centre of mass moves at constant speed. 
Equation \ref{rdotgr} needs to be solved implicitly for the pseudo-velocities $\mathbf{\tilde{v}_i}$, but this is easily accomplished iteratively, since the 1PN approximation is only appropriate when the perturbation is weak, and the physical and pseudo-velocities are approximately equal.
Operationally, if like in \reb, the integrator does not work in Jacobi coordinates, one simply calculates the Newtonian accelerations in inertial coordinates, transforms to Jacobi coordinates, calculates the Jacobi accelerations $\mathbf{\ddot{r}_i}$ and transforms back to inertial coordinates.
\cite{WHFAST} provide unbiased transformation algorithms between inertial and Jacobi coordinates.

\section{Centre of Mass} \label{COM}
When inserting additional effects in N-body simulations in an inertial frame, it is often valuable to ensure that there are no net forces acting on the system's centre of mass (COM).
For example, the Wisdom-Holman integration scheme \citep{Kinoshita90, Wisdom91, WHFAST} assumes that the COM moves at constant velocity and removes that degree of freedom from the problem.
This effectively means that any residual net force is incorrectly distributed among the particles' relative coordinates.
However, even integration schemes that track the COM degree of freedom can be adversely affected by finite forces on the barycentre.
Such a simulation's accuracy will continually deteriorate as the system moves away from the origin and the precision on the inter-particle separations degrades due to subtracting ever larger nearly equal numbers.
These errors may be acceptable depending on the application, but can slow down schemes like IAS15 \citep{IAS15} that try to reach machine precision as they fail to converge with progressively smaller timesteps.

It is therefore numerically desirable to self-consistently model all back-reactions from additional effects so that the net force vanishes.  
While this is trivial for forces between pairs of particles in the simulation, one does not always wish to fully model all components.

Consider a net force $\mathbf{F_D}$ from a protoplanetary disk of mass $M_D$ on a planet.  
In response, the disk should feel an equal and opposite net force, but there is no disk ``particle" in the simulation on which to apply it.
This uncompensated force will yield an acceleration on the COM.
However, the disk is tightly coupled to the central star gravitationally, suggesting that this back-reaction should be effectively communicated to the central body.
More quantitatively, if we take a characteristic orbital radius for the disk $r_D$, then we can define a characteristic timescale for the back-reaction $\tau_{BR} \sim (r_D/a_D)^{1/2}$, where $a_D$ is the net acceleration of the disk $F_D/M_D$.
At the same time, the disk is gravitationally coupled to the primary on the Keplerian orbital period $\tau_K = 2\pi (GM/r_D^3)^{-1/2}$.
If $\tau_K \ll \tau{BR}$, the central body will respond adiabatically to the accelerations of the disk, and the back-reaction could instead be added to the central star.

In other cases it might be better motivated to add the accelerations to the centre of mass of all or a subset of the particles, e.g., back-reactions onto a circumbinary disk.
This corresponds to applying an acceleration $\mathbf{a} = \mathbf{-F}/M_{tot}$ to the appropriate set of particles, where $M_{tot}$ is their total mass.

In the case of forces that only depend on particle positions, the above choices are straightforwardly reflected in the structure of the Hamiltonians that govern the dynamics.
Being able to precisely calculate these Hamiltonians provids a practical check on the numerical accuracy of the implementation for conservative systems.
A Hamiltonian that depends on the radial positions $\mathbf{r_i}$ and $\mathbf{r_j}$ of two particles, $\mathcal{H}(\mathbf{r_i} - \mathbf{r_j})$ exerts equal and opposite forces on the two bodies.
Similarly, one can show that for a Hamiltonian dependent on the distance from the system's barycentre $\mathcal{H}(\mathbf{r_i - r_{COM}})$, where $\mathbf{r_{COM}} = M_{tot}^{-1} \sum_j m_j \mathbf{r_j}$, Hamilton's equations dictate that if particle $i$ feels force $\mathbf{F_i} = -\nabla_{r_i} \mathcal{H}(\mathbf{r_i - r_{COM}})$, all particles (including $i$) should feel an additional acceleration $-\mathbf{a_{COM}} = -\mathbf{F_i}/M_{tot}$.
Finally, for a Hamiltonian with positions referenced to the centre of mass of a subset of particles, Hamilton's equations require that only those particles feel the reverse acceleration $-\mathbf{F_i}/M_{sub}$, where $M_{sub}$ is the mass of the subset of particles.
In the case where each particle's position is referenced to the centre of mass of all interior bodies, this corresponds to the familiar Jacobi coordinates.

\end{document}